\documentclass[12pt]{article}

\usepackage{amsmath}
\usepackage{amsfonts}
\usepackage{amsthm}
\usepackage{latexsym}
\usepackage{mathrsfs}

\renewcommand{\theequation}{\thesection.\arabic{equation}}

  \def\D{{\mathscr D}}
  \def\Db{\bar{\!\D\,}\!}

  \def\L{{\mathscr L}}

\def\e{{\rm e}}

\def\l{\langle}
\def\r{\rangle}
\def\pr{\partial}

\newcommand{\half}{{\textstyle \frac{1}{2}}}
\def\quar{{\textstyle \frac{1}{4}}}

\newcommand{\Geff}{\Gamma_{\rm eff}}

\newcommand{\dS}{\!\!{\rm d}^6z\,}
\newcommand{\dSb}{\!\!{\rm d}^6\bar z\,}
\newcommand{\dV}{\!\!{\rm d}^8z\,}
\newcommand{\dx}{\!\!{\rm d}^4x\,}

\newcommand{\Ikin}{I_{\rm kin}}
\newcommand{\Ig}{I_{g}}
\renewcommand{\Im}{I_{M}}
\newcommand{\Ixi}{I_{\xi}}
\newcommand{\Ii}{I_{1}}
\newcommand{\Iii}{I_{2}}

\newcommand{\Lkin}{\L_{\rm kin}}
\newcommand{\Lg}{\L_{g}}
\newcommand{\Lm}{\L_{M}}
\newcommand{\Lxi}{\L_{\xi}}
\newcommand{\Lxib}{\L_{\xi}'}
\newcommand{\Li}{\L_{1}}
\newcommand{\Lib}{\L_{1}'}
\newcommand{\Lii}{\L_{2}}
\newcommand{\Liib}{\L_{2}'}
\newcommand{\Lbox}{\L_{\Box}}

\newcommand{\ukin}{u_{\rm kin}}

\newcommand{\uxi}{u_{\xi}}
\newcommand{\uxib}{u_{\xi}'}
\newcommand{\ui}{u_{1}}
\newcommand{\uib}{u_{1}'}
\newcommand{\uii}{u_{2}}
\newcommand{\uiib}{u_{2}'}
\newcommand{\ubox}{u_{\Box}}

\newcommand{\nIkin}[1]{I_{\rm kin}^{(#1)}}
\newcommand{\nIxi}[1]{I_{\xi}^{(#1)}}

\newcommand{\nIi}[1]{I_{\rm Ia}^{(#1)}}
\newcommand{\nIii}[1]{I_{\rm IIa}^{(#1)}}
\newcommand{\nIv}[1]{I_{\rm IIb}^{(#1)}}
\newcommand{\nIvi}[1]{I_{\rm IIc}^{(#1)}}
\newcommand{\nIvii}[1]{I_{\rm Ib}^{(#1)}}

\newcommand{\nLkin}[1]{\L_{\rm kin}^{(#1)}}
\newcommand{\nLxi}[1]{\L_{\xi}^{(#1)}}
\newcommand{\nLxib}[1]{{\L_{\xi}'}^{(#1)}}

\newcommand{\nLi}[1]{\L_{\rm Ia}^{(#1)}}
\newcommand{\nLib}[1]{{\L_{\rm Ia}'\!}^{(#1)}}
\newcommand{\nLvii}[1]{\L_{\rm Ib}^{(#1)}}
\newcommand{\nLviib}[1]{{\L_{\rm Ib}'\!}^{(#1)}}
\newcommand{\nLic}[1]{{\L_{\rm Ia}''}^{(#1)}}
\newcommand{\nLii}[1]{\L_{\rm IIa}^{(#1)}}
\newcommand{\nLiib}[1]{{\L_{\rm IIa}'\!\!}^{(#1)}}
\newcommand{\nLvi}[1]{\L_{\rm IIc}^{(#1)}}
\newcommand{\nLvib}[1]{{\L_{\rm IIc}'\!\!}^{(#1)}}
\newcommand{\nLiic}[1]{{\L_{\rm IIa}''\!\!}^{(#1)}}
\newcommand{\nLv}[1]{\L_{\rm IIb}^{(#1)}}
\newcommand{\nLvb}[1]{{\L_{\rm IIb}'\!\!}^{(#1)}}
\newcommand{\nLvc}[1]{{\L_{\rm IIb}''\!\!}^{(#1)}}

\newcommand{\nukin}[1]{u_{\rm kin}^{(#1)}}
\newcommand{\nuxi}[1]{u_{\xi}^{(#1)}}
\newcommand{\vxi}[1]{v_{\xi}^{(#1)}}
\newcommand{\nuxib}[1]{{u_{\xi}'}^{(#1)}}

\newcommand{\nui}[1]{u_{\rm Ia}^{(#1)}}
\newcommand{\nuib}[1]{{u_{\rm Ia}'\!\!}^{(#1)}}
\newcommand{\nuic}[1]{{u_{\rm Ia}''}^{(#1)}}
\newcommand{\nuvii}[1]{u_{\rm Ib}^{(#1)}}
\newcommand{\nuviib}[1]{{u_{\rm Ib}'\!\!}^{(#1)}}
\newcommand{\vi}[1]{v_{\rm Ia}^{(#1)}}
\newcommand{\vvii}[1]{v_{\rm Ib}^{(#1)}}
\newcommand{\nuii}[1]{u_{\rm IIa}^{(#1)}}
\newcommand{\nuiib}[1]{{u_{\rm IIa}'\!\!}^{(#1)}}
\newcommand{\nuiic}[1]{{u_{\rm IIa}''\!\!}^{(#1)}}
\newcommand{\nuv}[1]{u_{\rm IIb}^{(#1)}}
\newcommand{\nuvb}[1]{{u_{\rm IIb}'\!\!}^{(#1)}}
\newcommand{\nuvc}[1]{{u_{\rm IIb}''\!\!}^{(#1)}}
\newcommand{\nuvi}[1]{u_{\rm IIc}^{(#1)}}
\newcommand{\nuvib}[1]{{u_{\rm IIc}'\!\!}^{(#1)}}

\newcommand{\vii}[1]{v_{\rm IIa}^{(#1)}}
\newcommand{\vvi}[1]{v_{\rm IIc}^{(#1)}}
\newcommand{\vv}[1]{v_{\rm IIb}^{(#1)}}

\newcommand{\al}{\alpha}
\def\da{{\dot\alpha}}
\def\be{\beta}
\def\db{{\dot\beta}}

\newcommand{\tfr}[2]{{\textstyle \frac{#1}{#2}}}

\newcommand{\fdq}[2]{\frac{\delta #1}{\delta #2}}

\def\ts{\textstyle}

\renewcommand{\i}{{\rm i}}

\newcommand{\tL}{\tilde L}

\newcommand{\w}[2]{{\ts {{\hfill #1} \atop {\hfill #2}}\!w}{}}
\newcommand{\wb}[2]{{\ts {{\hfill #1} \atop {\hfill #2}}\!\bar w}{}}
\newcommand{\W}[2]{{\ts {{\hfill #1} \atop {\hfill #2}}\!W}{}}

\newcommand{\geom}{{\rm geom}}

\newcommand{\Pc}{(\Db^2+R)}
\newcommand{\Pac}{(\D^2+\bar R)}

\newtheorem{theorem}{Theorem}

\newcommand{\mytable}[3]{\begin{table}
\rule{\textwidth}{1pt}\\
#1
\rule{\textwidth}{0.5pt}\\[-5ex]
\caption{#2}
\vspace*{-1.5ex}
\rule{\textwidth}{1pt}
\label{#3}
\end{table}
}

\newcommand{\tl}{\tilde l}

\newcommand{\T}{\mbox{\bf T}\,}

\begin{document}
\vspace*{-5mm}
\begin{flushright}
hep-th/9910153\\ NTZ 23/1999\\ October 1999
\end{flushright}

\thispagestyle{empty}

\vspace{1cm}
\begin{center}
{\Large 
 \bf Conformal Transformation Properties  }

{\Large \bf of the B-type Supercurrent}

\vspace{1cm}

{\parindent0cm
Christian Rupp\footnote{E-mail: Christian.Rupp@itp.uni-leipzig.de}
}

\vspace{0.7cm}

Institut f{\"u}r Theoretische Physik\\
Universit{\"a}t Leipzig\\
Augustusplatz 10/11\\
D - 04109 Leipzig\\
Germany
\end{center}

\vspace{1.5cm}

\centerline{\small \bf Abstract}\vspace*{-2mm} { \small \noindent 
We investigate the superconformal transformation properties of
the supercurrent as well as of the superconformal anomalies themselves in
$d=4$, $N=1$ supersymmetric quantum field theory. 
Matter supercurrent and anomalies are coupled to a classical background of
minimal supergravity fields. On flat superspace, there exist two different 
types of the superconformal Ward identity (called S and B) which correspond
to the flat space 
limits of old resp.\ new minimal background supergravity fields. In the present
publication we give particular importance to the new minimal case.
A general formalism is set up which is then applied to the massless
Wess-Zumino model.
}

\vspace*{15mm}
\begin{tabbing}
\> Curved Superspace
Background, New Minimal Supergravity, \\ \> Supercurrent, Anomalies.
\end{tabbing}

\newpage

\section{Introduction}
\setcounter{equation}{0}
\setcounter{table}{0}

It is well-known that on flat $d=4$, $N=1$ superspace there exist two different types (called S and B) of
superconformal symmetry breaking and correspondingly two versions of the
supercurrent \cite{PScurrent, PSBuch}. 
In a recent series of papers \cite{ERS1, ERS2, ERS3, ERS4}, the S-type supercurrent has been
studied by coupling it to a background of old minimal supergravity
fields, such that the superconformal Ward identity on flat space
corresponds to the flat space limit of a combined diffeomorphism and Weyl
Ward identity. By 
functionally differentiating with respect to the supergravity field
$H^{\al\da}$, the superconformal transformation properties of Green
functions with one or more insertions of the S-type supercurrent have been
explicitly determined to all orders in perturbation theory for the
Wess-Zumino model as well as for SQED. It has been
shown there that the anomalous breaking of Weyl symmetry on curved space
may be expressed 
in terms of a {\em local Callan-Symanzik equation}, in which all dynamical
anomalies are parametrized by the $\beta$ and $\gamma$ functions of the
theory. This equation already contains all information on superconformal
anomalies of multiple supercurrent insertions.

In the present article, the B-type supercurrent is considered in a similar
way by using a background of {\em new minimal} supergravity
fields. Furthermore, we also investigate the transformation properties of
the anomalous breaking terms $B_\al$ and $S$ themselves.

This article is organized as follows.
In section \ref{sec:SBWI}, superconformal Ward identities on flat space
are briefly reviewed, including the two possible types of breaking terms.
In section \ref{sec:classical}, the formalism of coupling the supercurrent
as well as the breaking terms to external fields is presented in a model
independent way. The types S and B of the superconformal Ward identity are
shown to correspond to the flat space limit of {\em old} resp.\ {\em new}
minimal supergravity backgrounds. 
The transition from the old to the new formulation is achieved by
introducing an additional external field $L$ (which is linear and real) in
a Weyl invariant way. The choice of parametrization of the new minimal
supergravity is non-trivial here, the primary requirement being the
possibility to vary all fields independently. This problem is solved by
expressing $L$ in terms of a flat space chiral spinor superfield $\eta_\al$.
By taking functional derivatives with
respect to the external supergravity fields, the transformation properties
of supercurrent and breaking terms are obtained. Since all Ward identities
are formulated off-shell and for arbitrary functionals, they may be applied
to the classical action as well as to the vertex functional.
In section \ref{sec:quant} we consider the massless Wess-Zumino model in a
perturbative approach using BPHZL renormalization. A close analysis of all
possible breaking terms shows that the Weyl invariant coupling of $L$ is
possible, such that the formalism of section \ref{sec:classical} may be
applied.

\section{Superconformal Ward identities}
\label{sec:SBWI}
\setcounter{equation}{0}
\setcounter{table}{0}

According to the Noether theorem, for each continuous symmetry there exists a
conserved current. At first, these currents are fields on Minkowski space
and are given in terms of components of the superfields involved. As usual,
the energy-momentum tensor corresponds to translational invariance, while supersymmetry
implies the existence of a conserved current called {\em supersymmetry
  current}. 
This supersymmetry current $Q_{a \al}$ should be clearly distinguished from the {\em
  supercurrent} $V_{\al\da}$ which is a superfield having the
energy-momentum tensor  $T_{ab}$,
supersymmetry current $Q_{a\al}$ and $R$-current $R_a$ amongst its components.

Suppose we have a superconformally invariant action $\Gamma$. Then the
supercurrent 
is itself conserved in the sense that the Ward identities
\begin{subequations}
\label{confWI}
\begin{align}
8\i w\Gamma &= \pr^a V_a
 \label{confWI1}\\[0.2cm]
-16 w_\al \Gamma &= \bar D^\da V_{\al\da}  \label{confWI2}\\[0.2cm]
-16 \bar w_\da \Gamma &= D^\al V_{\al\da} \label{confWI3}
\end{align}
\end{subequations}
hold,
where $w=D^\al w_\al - \bar D_\da \bar w^\da$, and $w_\al$ is the {\em local
  Ward operator} of superconformal transformations. If $\Gamma$ depends
  only on a chiral field $A$, $w_\al$ is given by
\begin{equation}
w_\al = \quar D_\al A \fdq{}{A} - \tfr{1}{12}D_\al
\left(A\fdq{}{A}\right)\,.
\label{firstward}
\end{equation}
The meaning of $w_\al$ has been discussed in detail in \cite{ERS1}. At this
point it is 
sufficient to note that $w_\al$ is a functional differential operator such
that $w_\al \Gamma$ vanishes on-shell. Thus the divergence of $V_a$ also
vanishes on-shell, and $V_a$ is a conserved current. Equations
(\ref{confWI2}), (\ref{confWI3}) represent a decomposition of
  (\ref{confWI1}) and are called {\em trace identities}. 

In the superconformal case, conserved R-current, energy-momentum tensor and supersymmetry
current are given by
\begin{subequations}
\label{RQT_Bdef}
\begin{align}
R_a &= \half C_a\\
Q_{a \al} &= \tfr{\i}{2} \chi_{a\al} \\
T_{ab} &= -\tfr{1}{4} v_{(ab)}\,,
\end{align}
\end{subequations}
where
\begin{equation}
V_a (z)= C_a(x) + \theta^\al \chi_{a\al}(x) + \bar \theta_\da \bar
\chi_a{}^\da (x) + \theta \sigma^b \bar \theta v_{ab}(x) + \dots.
\label{Vcomp}
\end{equation}

Let us now turn to the case of broken superconformal invariance. In this
case, the Ward identities (\ref{confWI}) contain additional breaking terms
$B_\al$ and $S$:
\begin{subequations}
\label{brconfWI}
\begin{align}
8\i w\Gamma &= \pr^a V_a
-\i \left( D^2 S - \bar D^2 \bar S \right)\,, \label{brconfWI1}\\[0.2cm]
-16 w_\al \Gamma &= \bar D^\da V_{\al\da}  -B_\al + 2 D_\al S\,, \label{brconfWI2}\\[0.2cm]
-16 \bar w_\da \Gamma &= D^\al V_{\al\da} - \bar B_\da  + 2 \bar D_\da \bar
S \,,\label{brconfWI3}
\end{align}
\end{subequations}
where
\begin{subequations}
\label{BSconstr}
\begin{gather}
\bar D_\da S = 0\,, \label{BSconstr1}\\
\bar D^2 B_\al = 0\,, \qquad D^\al B_\al - \bar D_\da \bar B^\da =0 \,.
\label{BSconstr2} 
\end{gather}
\end{subequations}
The restrictions (\ref{BSconstr}) on $B_\al$  and $S$ ensure that $\Gamma$
is still super Poincar\'e invariant. It has been shown however
\cite{PScurrent, PSBuch},
that a  conserved energy-momentum tensor and supersymmetry current can be formed from the
components of $V_a$ only if either $B_\al$ or $S$ vanishes\footnote{A
  conserved energy-momentum tensor and supersymmetry current do also exist if $B_\al$ and $S$
  are both different from zero. But in this case they can not be expressed
  in terms of components of the supercurrent only.}. The
decomposition (\ref{brconfWI2}) of $w_\al \Gamma$ into $V$, $S$ and $B$ is
far from being unique. Indeed, in all known cases it is possible to
eliminate $B_\al$ by a redefinition of $V$ and $S$. In some cases -- but
not always -- it is also possible to eliminate $S$ by a redefinition of $V$
and $B$. The case $S=0$ is denoted as {\em B formulation}, the
corresponding supercurrent as {\em B-type supercurrent}. Analogously, for
$B=0$ we have the {\em S formulation} and the {\em S-type supercurrent}.
In the superconformal case, both formulations coincide.

From (\ref{brconfWI1}) it is clear that if $S=0$ then (\ref{RQT_Bdef})
yields conserved currents. This implies that the B formulation can only be
possible for R-invariant theories.
If $S\ne 0$, $\Gamma$ may or may not be R-invariant. Furthermore $R$, $Q$
and $T$ defined by (\ref{RQT_Bdef}) are not conserved as (\ref{brconfWI1})
shows. There exists, however, a different definition of conserved currents
$Q_{a \al}$ and $T_{ab}$ in terms of supercurrent components
\cite{PScurrent},
given by
\begin{align}
T_{ab} &= -\tfr{1}{8} (v_{ab} + v_{ba} - 2 \eta_{ab} v^c{}_c)\,, 
\label{Tabexplicit}\\
Q_a{}^\al &= \i \left( \chi_a{}^\al - (\sigma_a \bar\sigma_b)_\al{}^\be
  \chi^b{}_\be \right)\,,\\            
R_a(x) &= \half C_a(x)\,.
\end{align}

\section{Classical Formalism}
\label{sec:classical}
\setcounter{equation}{0}
\setcounter{table}{0}

\subsection{Old minimal supergravity background}
\label{sec:oldmin}
The covariant derivatives 
\begin{gather}
\D_A \equiv ( \D_a, \D_\al, \bar \D^{\da} )
\end{gather}
of old minimal supergravity are determined by the prepotentials $H$ and
$\phi$, where $H=H^a\pr_a$ is a real vector superfield and $\phi=\e^J$ is
chiral. On flat space ($H=0$, $J=0$) they reduce to the usual derivatives
\begin{gather}
D_A \equiv ( \pr_a, D_\al, \bar D^{\da} ) \, .
\end{gather}
In this paper we use the real as well as the curved space chiral
representation. Quantities in the real representation are denoted by
letters with a tilde, e.g. $\tilde \Phi$. The corresponding expression in
the chiral representation is given by $\Phi = \e^{\i H} \tilde \Phi$.
From the fact that this transformation with $\e^{\i H}$ does not commute
with complex conjugation, there arises a notational difficulty. For
example, the complex conjugate $\D_\al$ of the chiral representation
covariant derivative $\bar D_\da$ is in the antichiral representation,
while in the chiral representation it reads
\begin{equation}
\e^{2\i H} D_\al \e^{-2\i H}\,.
\end{equation}

Superdiffeomorphisms are generated by vector fields
\begin{equation}
\Lambda=\Lambda^a \pr_a + \Lambda^\al D_\al + \Lambda_\da \bar D^\da +
\Lambda^{\al\be} M_{\al\be} + \Lambda^{\da\db} \bar M_{\da\db}
\end{equation}
which are subject to certain constraints, such that they may be expressed
in terms of a single parameter function $\Omega^\al(z)$ as
\begin{gather}
\Lambda^{\alpha\dot\alpha} = i \bar D^{\dot\alpha} \Omega^\alpha\, ,\qquad
\Lambda^\alpha = \tfr{1}{4} \bar D^2 \Omega^\alpha \, , \qquad
\Lambda_\da = \e^{2\i H} \bar \Lambda_\da\,, \\
\Lambda_{\da\db} = - \bar D_{(\da} \Lambda_{\db)}\,, \qquad 
\Lambda_{\al\be} = \e^{2\i H} \bar \Lambda_{\al\be}\,.
\label{omegadef}
\end{gather}
The diffeomorphism transformation properties of the supergravity
prepotentials are given by
\begin{equation}
\e^{2\i H} \longrightarrow \e^\Lambda \e^{2\i H} \e^{-\bar \Lambda}
\,,\qquad
\phi^3 \longrightarrow \phi^3 \e^{\overleftarrow{\Lambda_c}}\,,
\end{equation}
where $\Lambda_c = \Lambda^a\pr_a + \Lambda^\al D_\al$.
Super Weyl transformations are local scalings of the chiral compensator
$\phi$ but leave $H$ invariant,
\begin{equation}
\phi \longrightarrow \e^\sigma \phi\,, \qquad H \longrightarrow H\,.
\end{equation}
Superconformal transformations are now conveniently characterized as those
combined $\Lambda$ and Weyl transformations which leave the flat space
choice $H=0$, $J=0$ invariant, i.\ e.\
\begin{equation}
\sigma = -\tfr{1}{12} \bar D^2 D^\al \Omega_\al \,, \qquad
\bar D^\da \Omega^\al = D^\al \bar \Omega^\da\,. \label{confconstr}
\end{equation}

Transformation properties of functionals such as the classical action or the
vertex functional are most adequately expressed in terms of Ward
identities. Local Ward operators are denoted by $w$, surrounded by some
indices. The upper left index contains the fields which are to be
transformed, e.\ g. $\w{H}{}$ or $\w{J}{}$ or $\w{\rm dyn}{}$ for the
dynamical fields (as opposed to the background fields). 
The type of transformation is specified in the lower left position:
$\w{}{\Lambda}$ for diffeomorphisms, $\w{}{\sigma}$ for Weyl
transformations and $\w{}{\Lambda\sigma}$ for combined $\Lambda$ and Weyl
transformations. Integrated Ward operators are denoted by $W$ and are
defined by
\begin{equation}
\W{H}{\Lambda} \Gamma = \int \dV \Omega^\al(z) \w{H}{\Lambda}_\al(z) \Gamma +
c.c. = \int \dV \delta_\Lambda H^{\al\da} \, \fdq{\Gamma}{H^{\al\da}} + c.c.\,.
\end{equation}
Thus diffeomorphism invariance of the action $\Gamma$ is
expressed by
\begin{equation}
\left( \w{H}{\Lambda}_\al + \w{J}{\Lambda}_\al + \w{\rm dyn}{\Lambda}_\al
\right) \Gamma =0\,. \label{WIdiff}
\end{equation}
For Weyl
transformations we include a general chiral breaking term $S$,
\begin{equation}
\left( \w{J}{\sigma} + \w{\rm dyn}{\sigma} \right) \Gamma = - \tfr{3}{2}
S\,. \label{WIweyl}
\end{equation}
We may now combine Ward identities (\ref{WIdiff}) and (\ref{WIweyl}) by
imposing the first of constraints (\ref{confconstr}), such that the
inhomogeneous terms in the $\Lambda$ and Weyl transformation of $J$  cancel. 
This means that in the flat space limit there is no contribution from $J$.
However, the inhomogeneous term in the local Ward operator for $H$,
\begin{equation}
\w{H}{\Lambda}_\al^{\rm inhom} = \half \bar D^\da \fdq{}{H^{\al\da}}\,,
\end{equation}
is also present on flat superspace. We obtain
\begin{equation}
\w{\rm dyn}{\Lambda\sigma}_\al \Gamma = - \half \bar D^\da
\fdq{\Gamma}{H^{\al\da}} - \tfr{1}{8} D_\al S\,,
\end{equation}
which is exactly the S-type superconformal Ward identity since $V_{\al\da}
= 8 \fdq{\Gamma}{H^{\al\da}}$.
By differentiating once with respect to $H^{\al\da}$ before restricting to
flat space and by integrating with the parameter function $\Omega_{\rm
  conf}^\al(z)$ for conformal transformations, one
obtains the transformation properties of the supercurrent (see \cite{ERS1}
for details), 
\begin{align}
\W{\rm dyn}{\Lambda\sigma}(\Omega_{\rm conf}) V_{\al\da}(z) & = \delta
V_{\al\da}(z) - \tfr{3}{2} \int \dS' \sigma(z') \fdq{S(z')}{H^{\al\da}(z)}
- \tfr{3}{2} \int \dSb' \bar \sigma(z') \fdq{\bar
  S(z')}{H^{\al\da}(z)}\,,\\
\delta V_{\al\da} &= \Lambda V_{\al\da} - \tfr{3}{2} (\sigma+\bar \sigma)
V_{\al\da}\,,\\
\W{\rm dyn}{\Lambda\sigma}(\Omega) &= \int \dV \left(\Omega^\al (z) \w{\rm
  dyn}{\Lambda\sigma}_\al(z) + \bar \Omega_\da(z) \wb{\rm
  dyn}{\Lambda\sigma}^\da(z) \right)\,.
\end{align}
Here we have assumed that the dynamical fields transform independently of
$H$, as is the case for chiral scalar fields. Otherwise additional terms
may be present. $\Omega_{\rm conf}$ and $\sigma$ are the solutions of
(\ref{confconstr}).

Next we aim at a curved space extension of the B formulation of the
superconformal Ward identity. It turns out that this is achieved by using
the curved space formalism of {\em new minimal supergravity} which is
obtained from the {\em old minimal} formulation by introducing an
additional external field $L$ which is linear and real. We proceed in two
steps. In the first step, a real external field without the linearity
constraint is coupled to the breaking term $S$. As a by-product, this
allows the calculation of the conformal transformation properties of $S$
itself. In the second step, the linearity condition is employed. Since
functional derivatives with respect to constrained fields are not
well-defined, the linear field $L$ is expressed in terms of a flat space
chiral spinor field which turns out to couple directly to the $B$ breaking.

\subsection{Coupling of the Weyl breaking to an external field}
\label{sec:weylcoupling}

We introduce an additional external field $L$ which is restricted to be real.
In the chiral representation this means
\begin{equation}
L=\e^{2\i H} \bar L \,,
\end{equation}
thus $L$ is subject to an $H$-dependent constraint and cannot be varied
independently from $H$. It is therfore preferable to transform $L$ back
from the chiral to the real representation, i.e. to use 
\begin{equation}
\tilde L = \e^{-\i H} L
\end{equation}
instead of $L$ as independent field. $\tilde L$ is a real field in the
usual sense,
\[ \tilde L = \bar{\tilde L}\,. \]
By definition $L$ transforms under $\Lambda$ transformations as a scalar field,
\begin{equation}
\delta_\Lambda L = \Lambda L\,,
\end{equation}
which yields together with the transformation of $H$ (\ref{delta_Omega_H}) the
transformation of $\tilde L$:
\begin{subequations}
\label{LambdaLtilde}
\begin{align}
\delta_\Lambda \tilde L &=\delta_\Lambda^{(0)} \tilde L  +
\delta_\Lambda^{(1)} \tilde L + O(H^2) \\
\delta_\Lambda^{(0)} \tilde L &= \tfr{1}{4} \bar D^\da \Omega^\al \{ D_\al,
\bar D_\da \} \tilde L  + \tfr{1}{4} \bar D^2 \Omega^\al D_\al \tilde L +
c.c. \\
\delta_\Lambda^{(1)} \tilde L &= -\tfr{1}{16} H^{\al\da} \{ D_\al, \bar
D_\da \} \bar D^\db \Omega^\be \{ D_\be, \bar D_\db \} \tilde L 
- \tfr{1}{8} H^{\al\da} \{ D_\al, \bar D_\da \} \bar D^2 \Omega^\be D_\be
\tilde L \nonumber  \\
& \quad + \tfr{1}{16} \{ D_\al, \bar D_\da\} H^{\be\db} \bar D^\da
\Omega^\al \{ D_\be, \bar D_\db \} \tilde L + c.c.
\end{align}
\end{subequations}
The Weyl transformation of $\tilde L$ is defined as
\begin{equation}
\delta_\sigma \tilde L = - \left( \e^{-\i H} \sigma + \e^{\i H} \bar \sigma
\right) \tilde L  \,.
\label{WeylLtilde}
\end{equation}
From (\ref{LambdaLtilde}) and (\ref{WeylLtilde}) the Ward operators may be
calculated 
\begin{align}
\w{\tilde L}{\Lambda}_\al  &= \w{\tilde L}{\Lambda}_\al^{(0)}+  \w{\tilde
  L}{\Lambda}_\al^{(1)}+ O(H^2) \label{tLWard1}\\
\w{\tilde L}{\Lambda}_\al^{(0)} &= \tfr{1}{4} \bar D^\da \left( \{ D_\al,
  \bar D_\da \} \tilde L \fdq{}{\tL} \right) + \tfr{1}{4} \bar D^2 \left(
  D_\alpha \tL \fdq{}{\tL} \right) \\
\w{\tL}{\Lambda}_\al^{(1)} &= \tfr{1}{16} \{ D_\be, \bar D_\db \} \bar
  D^\da \left( H^{\be\db} \{ D_\al, \bar D_\da \} \tL \fdq{}{\tL} \right)
+ \tfr{1}{8} \{ D_\be, \bar D_\db\} \bar D^2 \left( H^{\be\db} D_\al \tL
  \fdq{}{\tL} \right) \nonumber \\
& \quad + \tfr{1}{16} \bar D^\da \left( \{ D_\al, \bar D_\da \} H^{\be\db}
  \{ D_\be, \bar D_\db \} \tL \fdq{}{\tL} \right) + c.c. \\
\w{\tL}{\sigma} &= - \bar D^2 \left( \left( \tL \fdq{}{\tL} \right) \e^{\i
  \overleftarrow{H}} \right) \label{tLWard4}
\end{align}

The new field $\tL$ is useful if it can be coupled in such a way that the
theory under consideration becomes Weyl invariant up to purely geometrical
terms. This is not always possible, a necessesary condition being $R$
invariance of the original theory. 
Futhermore $\tL$ should be coupled such that for $\tL=1$ we come back to
the original theory.

Let's assume we have succeeded in introducing $\tL$ as requested. Then the
following Ward identities hold:
\begin{subequations}
\label{LWI}
\begin{align}
\left( \w{\rm dyn}{\Lambda}_\al+\w{HJ\tL}{\Lambda}_\al\right) \,\, \Gamma
&= 0  \label{LWIa}\\
\left( \w{\rm dyn}{\sigma}+\w{J\tL}{\sigma}\right) \,\, \Gamma &=
-\tfr{3}{2} S_\geom(H,J,\tL)\,.
\label{LWIb}
\end{align}
\end{subequations}
For $\tL=1$ this implies
\begin{align}
\left( \w{\rm dyn}{\Lambda}_\al+\w{HJ}{\Lambda}_\al\right) \,\, \Gamma &= 0 \nonumber \\
\left( \w{\rm dyn}{\sigma}+\w{J}{\sigma} \right) \,\, \Gamma &= -\tfr{3}{2}
S_\geom(H,J) + \left. \bar D^2 \left( \fdq{\Gamma}{\tL} \e^{\i
    \overleftarrow{H}} \right) \right|_{\tL=1}\,.
\end{align}
This means that the dynamical anomalies of the original theory are now
coupled to $\tL$,
\begin{equation}
S_{\rm dyn}({\rm dyn}, H, J) =- \tfr{2}{3} \left. \bar D^2 \left( \fdq{\Gamma}{\tL}
  \e^{\i \overleftarrow{H}} \right) \right|_{\tL=1}\,.
\label{SLcoupling}
\end{equation}
Since the Weyl breaking is now controlled by $\tL$ we may confidently put
$J=0$ and accept the fusion of $\Lambda$ and Weyl transformations according
to (\ref{confconstr}),
\begin{equation}
\left( \w{\rm dyn}{\Lambda\sigma}_\al + \w{H\tL}{\Lambda\sigma}_\al\right) \,
\Gamma =  -\tfr{1}{8} D_\al S_\geom(H,J,\tL) \,.
\end{equation}
On flat space ($H=0$, $\tL=1$), the Ward identity reduces to
\begin{equation}
\w{\rm dyn}{\Lambda\sigma}_\al \, \Gamma = -\tfr{1}{2} \bar D^\da
\fdq{\Gamma}{H^{\al\da}} +\tfr{1}{12} D_\al \bar
D^2 \fdq{\Gamma}{\tL}\,.
\end{equation}

\subsection[Conformal transformation properties of the $S$ breaking]{Conformal
  transformation properties of the \boldmath $S$ \unboldmath breaking}

Since flat space is restored at $\tilde L=1$, it is convenient to use
$\tilde l \equiv \log \tilde L$ instead of $\tilde L$, such that flat space
corresponds to $\tilde l=0$. The transformations of $\tilde l$ are given by
\begin{align}
\delta_\Lambda^{(0)} \tilde l &= \tfr{1}{4} \bar D^\da \Omega^\al \{ D_\al,
\bar D_\da \} \tilde l  + \tfr{1}{4} \bar D^2 \Omega^\al D_\al \tilde l +
c.c. \\
\delta_\sigma \tilde l &= - \left( \e^{-\i H} \sigma + \e^{\i H} \bar \sigma
\right)
\end{align}
(\ref{SLcoupling}) translates into
\begin{equation}
S_{\rm dyn}({\rm dyn}, H, J) =- \tfr{2}{3} \left. \bar D^2 \left( \fdq{\Gamma}{\tl}
  \e^{\i \overleftarrow{H}} \right) \right|_{\tl=0}\,.
\label{Slcoupling}
\end{equation}
Ward operators for $\tl$ may simply be obtained from (\ref{tLWard1}) -
(\ref{tLWard4}) by noting that $\fdq{}{\tL}=\tL^{-1} \fdq{}{\tl}$.
By rewriting (\ref{LWI}) in terms of $\tl$, forming the conformal Ward
operator $\W{}{\Lambda\sigma}(\Omega_{\rm conf})$ and differentiating with
respect to $\tl$, one finds the flat space transformation law of the $S$
breaking.

\begin{theorem}
\label{thm_Strans}
In a model which is coupled to a real superfield $L=\e^{\i H}\e^{\tl}$ in a
Weyl invariant way, the following Ward identities hold at $H=0$, $\phi=1$,
$\tl=0$: 
\begin{align}
\w{\rm dyn}{\Lambda\sigma}_\al \Gamma &= -\tfr{1}{2} \bar D^\da
\fdq{\Gamma}{H^{\al\da}} +\tfr{1}{12} D_\al \bar D^2 \fdq{\Gamma}{\tl} 
\label{SWIl}\\[0.3cm]
\W{\rm dyn}{\Lambda\sigma}(\Omega_{\rm conf}) \bar D^2 \fdq{\Gamma}{\tilde l} &=
(\Lambda_c-3\sigma) \bar D^2 \fdq{\Gamma}{\tilde l} \nonumber \\
&\quad + 
\int \dS' \sigma(z') \bar D'{}^2 \bar D^2 \fdq{^2\Gamma}{\tilde l(z) \delta
  \tilde l(z')}
+\int \dSb' \bar\sigma(z') {D'}^2 \bar D^2 \fdq{^2\Gamma}{\tilde l(z) \delta
  \tilde l(z')}
\label{Stransclass}
\end{align}
\end{theorem}
One might expect a contribution to (\ref{Stransclass}) from the geometrical
anomalies in (\ref{LWIb}). The only chiral term of dimension $3$ containing
only one field $\tl$ is given by
\begin{equation}
S_{\rm geom}(H,J,\tl) = c \bar D^2 \Box \tl + O(\tl^2) + O(H) +O(J)\,.
\end{equation}
Thus there is no contribution to (\ref{Stransclass}) since
\begin{equation}
\int \dS \sigma c \bar D^2 \Box \tl = c\, \int \dV \Box \sigma \, \tl =0\,.
\end{equation}

\subsection{B formulation}
A preliminary version of the B formulation may be obtained already at this
point. 
We use the relation
\[D_\al \bar D^2 = 2\bar D^\da [D_\al, \bar D_\da] - 3 \bar D^2 D_\al
\]
to rewrite $S_{\rm dyn}$ as
\begin{equation}
\tfr{1}{8} D_\al S_{\rm dyn}({\rm dyn},H=0) 
= \tfr{1}{2} \bar D^\da \left( \tfr{1}{3} [D_\al, \bar D_\da]
 \fdq{\Gamma}{\tl} \right) - \tfr{1}{4} \bar D^2 D_\al \fdq{\Gamma}{\tl} \,.
\label{B1}
\end{equation}
It seems obvious to interpret the first term on the right hand side of
(\ref{B1}) as a contribution to the supercurrent while the second term has
the properties of a $B$ breaking:
\begin{align}
V^B_{\al\da} &= 8 \left( \fdq{}{H^{\al\da}} - \tfr{1}{3} [D_\al, \bar D_\da]
  \fdq{}{\tl} \right) \, \Gamma 
\label{VBLdef}\\
B_\al &= -4 \bar D^2 D_\al \fdq{\Gamma}{\tl} \,.
\label{BLdef}
\end{align} 
Using these definitions, (\ref{SWIl}) becomes 
\begin{equation}
\w{\rm dyn}{\Lambda\sigma}_\al \Gamma = -\tfr{1}{16} \bar D^\da
V^B_{\al\da} + \tfr{1}{16} B_\al\,.
\end{equation}
Furthermore, $B_\al$ as given by (\ref{BLdef}) obeys the constraint
(\ref{BSconstr}).

\subsection{Example: Classical O'Raifeartaigh model}
As a simple example we consider the classical O'Raifeartaigh model without
spontaneous symmetry breaking. This model describes three chiral fields
$A_0$, $A_1$, $A_2$, the flat space action being given by
\begin{equation}
\Gamma = \tfr{1}{16} \sum_{k=0}^2\int\dV A_k \bar A_k + \left( \int \dS \left(
    \tfr{m}{4} A_1A_2 + \tfr{g}{32} A_0 A_1^2\right) + c.c. \right) \,.
\end{equation}
On (old minimal) curved superspace this becomes
\begin{equation}
\Gamma = \tfr{1}{16} \int\dV E^{-1} A_k \e^{2\i H} \bar A_k + \left( \int
    \dS \phi^3  \left( 
    \tfr{m}{4} A_1A_2 + \tfr{g}{32} A_0 A_1^2\right) + c.c. \right)\,.
\end{equation}

The Weyl transformation of the dynamical fields is defined as
\begin{equation}
A_0 \to \e^{-3\sigma} A_0\,, \qquad A_1 \to A_1\,, \qquad A_2 \to \e^{-3\sigma}
A_2\,. 
\end{equation}
The Weyl weights are chosen to fit the non-conformal $R$-weights
$n_0=n_2=-2$, $n_1=0$ of $A_0$, $A_2$ and $A_1$ (see \cite{PSBuch}). Our
formalism automatically incorporates general superconformal transformations
with dilatational weights $d_i=-\tfr{3}{2} n_i$ different from the
canonical dimension $1$ of the fields.

Now we introduce the real superfield $\tL$. Obviously
\begin{align}
\Gamma &= \tfr{1}{16} \int\dV E^{-1} \left( A_0 \e^{2\i H} \bar A_0 L^{-2} 
+ A_1 \e^{2\i H} \bar A_1 L +A_2 \e^{2\i H} \bar A_2 L^{-2} \right)
\nonumber \\
& \quad + \left(
    \int \dS  \left( 
    \tfr{m}{4} A_1A_2 + \tfr{g}{32} A_0 A_1^2\right) + c.c. 
\right)
\end{align}
is Weyl invariant. The field $L=\e^{\i H}\tL$ has been used in the chiral
representation as an abbreviation. However it shall be understood that
$\tl$ is viewed as independent field. Furthermore we have put $J=0$.
Now the Ward identity
\begin{align}
\w{A_k}{\Lambda\sigma} \Gamma &= -\tfr{1}{16} \bar D^\da V^B_{\al\da} +\tfr{1}{16} B_\al
\\
\intertext{holds, and $V^B_{\al\da}$ and $B_\al$ may be calculated using
  (\ref{VBLdef}), (\ref{BLdef}),}
V^B_{\al\da} &= -\tfr{1}{2} \sum_k (n_k+1) D_\al A_k \bar D_\da \bar A_k +
\tfr{1}{4} \sum_k n_k ( -A_k D_\al \bar D_\da \bar A_k + \bar A_k \bar
D_\da D_\al A_k) \\
B_\al &= \tfr{1}{4} \bar D^2 D_\al \left( 2 A_0 \bar A_0 - A_1 \bar A_1 +
  2 A_2 \bar A_2 \right)\,.
\end{align}
This coincides with the known $B$ type Ward identity in \cite{PSBuch}.

The O'Raifeartaigh model is also well suited to check
(\ref{Stransclass}). The $S$ breaking is given by
\begin{equation}
S = -\tfr{2}{3} \left. \bar D^2 \fdq{\Gamma}{\tl} \right|_{\tl=0} = -\tfr{1}{24} \bar D^2 \left(
  -2 A_0 \bar A_0 + A_1 \bar A_1 - 2 A_2 \bar A_2 \right)\,.
\label{SORaif}
\end{equation}
According to (\ref{Stransclass}), $S$ transforms as
\begin{equation}
\W{A_k}{\Lambda\sigma}(\Omega_{\rm conf}) S = \Lambda_c S -\tfr{1}{24} \bar
D^2 \left( (-5\sigma-2\bar \sigma) (-2 A_0 \bar A_0- 2 A_2 \bar A_2) +
  (-2\sigma+\bar\sigma) ( A_1 \bar A_1) \right)\,,
\end{equation}
which may be independently checked by applying the transformation laws of
$A_0$, $A_1$, $A_2$ on (\ref{SORaif}).

\subsection{B formulation and new minimal supergravity}
\label{sect_finalB}
So far the B formulation has the disadvantage that the B-type supercurrent
doesn't couple directly to $H^{\al\da}$ but is rather given by a
combination of functional derivatives with respect to $H^{\al\da}$ and
$\tl$ (\ref{VBLdef}). Even worse, $B_\al$ depends only on some components of
$\fdq{\Gamma}{\tl}$ while other components are actually not needed.
Both drawbacks may be fixed by restricting $L$ to be a real linear
superfield. By imposing this restriction we end up in a background of new
minimal supergravity. For the formalism of new minimal supergravity, see
\cite{newmin, Grisaru, Buchbinder}.

There is however the technical complication that it is not
possible to functionally differentiate with respect to a linear superfield.
Thus the linearity constraint has to be solved by expressing $L$ in terms
of unconstrained fields. As a first step, $L$ is expressed in terms of a
real flat linear superfield $L_0$. Again, we cast $L$ into the real
representation $\tilde L = \e^{-\i H}L$. We have to solve the equation (for
$J=0$) 
\begin{equation}
0 = \Pc \e^{\i H} \tilde L = \bar D^2 \left( E^{-1} \e^{\i H} \tilde L
\right)\,.
\end{equation}
One might be tempted to identify $L_0$ with the term in brackets on the
right hand side of this equation, however this expression is not real.
A real expression for $L_0$ is obtained by the following manipulations,
\begin{align}
0 &= \bar D^2 \left( E^{-1} \e^{\i H} \tilde L
\right) = \bar D^2 \left( (\tilde E^{-1}\tilde L)\, \e^{\i \overleftarrow H}
\right) \nonumber \\
&= \half \bar D^2 \left( \sum_{n=0}^\infty \tfr{1}{n!} \{ D^\al, \bar D^\da
  \} \left( H_{\al\da} (\tilde E^{-1} \tilde L) (\i\overleftarrow H)^{n-1}
  \right) \right) \nonumber \\
&= \bar D^2 \left(  (\tilde E^{-1} \tilde L) \cosh(\i \overleftarrow H) +
  \half \left(  (\tilde E^{-1} \tilde L)\frac{\sinh (\i \overleftarrow
      H)}{\i \overleftarrow H} H^{\al\da} \right) \overleftarrow{[D_\al,
  \bar D_\da ]} \right)\,,
\label{L0def1}
\end{align}
where $\tilde E^{-1}=E^{-1} \e^{-\i \overleftarrow H}$ is the inverse
vierbein determinant in the real representation. $\cosh (\i
\overleftarrow{H})$ is real because $\cosh (x)$ contains only even powers
of $x$. The same is true for $\sinh(x)/x$. Since furthermore $\tilde E$,
$\tL$ and $[D_\al, \bar D_\da]$ are real, a suitable definition of $L_0$ is
\begin{equation}
L_0 \equiv (\tilde E^{-1} \tilde L) \left(  \cosh(\i \overleftarrow H) +
  \half \frac{\sinh (\i \overleftarrow
      H)}{\i \overleftarrow H} H^{\al\da} \circ \overleftarrow{[D_\al,
  \bar D_\da ]} \right)\,.
\label{L0def}
\end{equation}
$L_0$ is restricted by
\begin{equation}
L_0 = \bar L_0 \,, \qquad \bar D^2 L_0 =0\,.
\end{equation}
The second step is to solve this constraint by
\begin{equation}
L_0 = \half \left( D^\al \eta_\al + \bar D_\da \bar \eta^\da \right)\,,
\label{etadef}
\end{equation}
where $\eta_\al$ is a flat chiral spinor superfield, $\bar D_\da
\eta_\al=0$.
Thus the real, curved space linear superfield $L$ is now expressed in terms
of $H$ and the flat space chiral spinor fields $\eta_\al$, $\bar\eta_\da$,
\begin{equation}
L = L(H, \eta, \bar \eta)\,.
\end{equation}
$\eta_\al$, $\bar \eta_\da$ are not subject to any further constraints and
are thus well suited as independent fields to formulate Ward identities
with\footnote{Chiral superfields are essentially unconstrained because
  they live on the smaller superspace with integration measure \quad $\dS={\rm
    d}^4\!x \,{\rm d}^2\!\theta$.}.

By expressing $L_0$ in terms of $\eta$, $\bar \eta$,
an additional gauge invariance has occurred: The replacement
\begin{equation}
\eta_\al \to \eta_\al + \i \bar D^2 D_\al K\,,\qquad \bar \eta_\da \to \bar
\eta_\da - \i D^2 \bar D_\da K \qquad \mbox{with } K=\bar K
\label{etagauge}
\end{equation}
leaves $L_0$ invariant. We denote this invariance as {\em $K$-gauge
  invariance}.

To further evaluate the connection between $\tL$ and $L_0$, we decompose
the differential operator which acts on $\tL$ in (\ref{L0def}) 
corresponding to the order in $H$,
\begin{equation}
L_0 = \sum_{n=0}^\infty Y^{(n)} \, \tL \qquad (Y^{(0)}=1)\,.
\end{equation}
The inverse operator is written as
\begin{equation}
\tL = \sum_{n=0}^\infty X^{(n)} \, L_0 \qquad (X^{(0)}=1)\,.
\end{equation}
The connection between $Y^{(n)}$ and $X^{(n)}$ is given by
\begin{align}
Y^{(1)} &= - X^{(1)} \nonumber \\
Y^{(2)} &= -X^{(2)} + X^{(1)}X^{(1)} \nonumber \\[-0.2cm]
& \,\,\,\vdots \nonumber
\end{align}
By expanding (\ref{L0def}) in a power series in $H$, one finds
\begin{align}
X^{(1)} L_0 &= \tfr{1}{6} [D_\al, \bar D_\da] H^{\al\da} L_0 - \half
[D_\al, \bar D_\da] (H^{\al\da} L_0)  \label{X_1}\\[0.3cm]
X^{(2)}L_0 &=  - \quar \{ D_\al, \bar D_\da \} \left( H^{\al\da} \{ D_\be, \bar D_\db \}
  H^{\be\db} L_0 \right) 
+ \tfr{1}{9} \{ D_\al, \bar D_\da \} H^{\al\da} \{
D_\be, \bar D_\db\} H^{\be\db} L_0  \nonumber\\
& \quad + \tfr{1}{6} \bar D_\db D_\be H^{\al\da}
D_\al \bar D_\da H^{\be\db} L_0 
 - \tfr{1}{8} H^{\al\da} \{ D_\al, \bar D_\da\} H^{\be\db} \{ D_\be, \bar
D_\db\} L_0 \nonumber\\
& \quad  -\tfr{1}{8} H^{\al\da} H^{\be\db} \{ D_\al, \bar D_\da\} \{
D_\be, \bar D_\db\} L_0
-\tfr{1}{6} \bar D_\db H^{\al\da} \{ D_\al, \bar D_\da\} D_\be H^{\be\db}
L_0 \nonumber \\
& \quad  -\tfr{1}{6} D_\be H^{\al\da} \{ D_\al, \bar D_\da\} \bar D_\db H^{\be\db}
L_0
+ \quar [D_\al, \bar D_\da] \left( H^{\al\da} [D_\be, \bar D_\db]
  (H^{\be\db} L_0) \right)  \nonumber\\
& \quad  + \tfr{1}{36} [D_\al, \bar D_\da] H^{\al\da} [D_\be, \bar D_\db]
H^{\be\db} L_0
-\tfr{1}{12} [D_\al, \bar D_\da] H^{\al\da} [D_\be, \bar D_\db] (H^{\be\db}
L_0 ) \nonumber \\
& \quad  + \tfr{1}{18} \bar D_\da D_\al H^{\al\da} D_\be \bar D_\db H^{\be\db}
L_0\,.
\end{align}
(\ref{X_1}) coincides with a first-order field redefinition suggested in
\cite{Grisaru}. Furthermore, (\ref{L0def}) is similar but not identical to
a useful field redefinition in abelian gauge theory \cite{ERS4, Grisaru}. 

Next, the transformation properties of $L_0$ have to be
determined. Clearly,
\begin{equation}
\delta_{\Lambda\sigma} L_0 = \sum_{n=1}^{\infty} \W{H}{\Lambda}\left(Y^{(n)}\right)\,
\sum_{m=0}^{\infty} X^{(m)} L_0 \, + \, \sum_{n=0}^{\infty} Y^{(n)} \left(
  \delta_{\Lambda\sigma}\tL \right)_{\tL=\sum\limits_m^{} X^{(m)}L_0}\,,
\label{L0trans}
\end{equation}
with $\delta_{\Lambda\sigma}\tL$ given by the sum of (\ref{LambdaLtilde})
and (\ref{WeylLtilde}).
To zeroth order in $H$ this yields
\begin{align}
\delta_{\Lambda\sigma}^{(0)} L_0 &= \W{H}{\Lambda}^{(0)}\left(Y^{(1)}\right)\, L_0 +
\left(\delta^{(0)}_{\Lambda\sigma}\tL \right)_{\tL=L_0} \nonumber \\
&= -\quar D^\al \bar D^2 \left( \Omega_\al L_0 \right)
 -\quar \bar D_\da  D^2 \left( \bar \Omega^\da L_0 \right)\,.
\label{deltaL00}
\end{align}
To first order in $H$, (\ref{L0trans}) yields
\begin{align}
\delta_{\Lambda\sigma}^{(1)} L_0 &= \W{H}{\Lambda}^{(0)}
\left(Y^{(1)}\right)\, X^{(1)} L_0 \,+\,
\W{H}{\Lambda}^{(1)}\left(Y^{(1)}\right) \, L_0 \,+\,
\W{H}{\Lambda}^{(0)}\left( Y^{(2)}\right) \, L_0 \nonumber \\
& \quad + \left(\delta_{\Lambda\sigma}^{(0)} \tL \right)_{\tL=X^{(1)}L_0}
\,
+ \, \left(\delta_{\Lambda\sigma}^{(1)}\tL \right)_{\tL=L_0} \, + \,
Y^{(1)}\, \left(\delta_{\Lambda\sigma}^{(0)} \tL \right)_{\tL=L_0}\,.
\end{align}
The explicit expression for $\delta^{(1)}_{\Lambda\sigma}L_0$ is lengthy
and is therefore not displayed. 

Since it is not possible to functionally
differentiate with respect to a linear field, Ward identities cannot be
formulated in terms of $L_0$. Instead, $\eta_\al$ has to be used. In order
to find the transformation properties of $\eta_\al$,
$\delta_{\Lambda\sigma}L_0$ has to be rewritten as
\begin{equation}
\delta_{\Lambda\sigma} L_0 = \half \left( D^\al \left( \delta_{\Lambda\sigma}
  \eta_\al\right)  + \bar D_\da \left(\delta_{\Lambda\sigma}\bar \eta^\da\right)  \right) \,.
\end{equation}
$\delta^{(0)}_{\Lambda\sigma} L_0$ in (\ref{deltaL00}) already has this
form, and it follows
\begin{equation}
\delta^{(0)}_{\Lambda\sigma} \eta_\al = -\half \bar D^2 (\Omega_\al L_0)\,,
\qquad
\delta^{(0)}_{\Lambda\sigma} \bar \eta_\da = -\half D^2 (\bar \Omega_\da
L_0)\,,
\end{equation}
where $L_0$ is given by (\ref{etadef}).
For $\delta_{\Lambda\sigma}^{(1)}\eta_\al$ we content ourselves with the
inhomogeneous part, i.e. the transformation at $L_0=1$. After tedious
calculations one finds
\begin{align}
\left.\delta^{(1)}_{\Lambda\sigma}\eta_\al\right|_{L_0=1}&= \bar D^2 \left( -\tfr{1}{4} \bar D_\db D_\be
  H^{\be\db} \Omega_\al + \tfr{1}{4} H^{\be\db} D_\be \bar D_\db
  \Omega_\al 
  - \tfr{1}{8} H_{\al\db} D^2 \bar \Omega^\db + \tfr{1}{4} D_\be H^{\be\db}
  D_\al \bar \Omega_\db \right) \,.
\label{deltaetaeins}
\end{align}

If we start from an action functional in which $L$ is coupled in a Weyl
invariant way (up to geometrical breaking terms), Weyl invariance is of
course still present when $L$ is expressed in terms of $\eta_\al$, i.e. the
following Ward identity holds:
\begin{equation}
\left( \w{\rm dyn}{\Lambda\sigma}_\al (z) + \w{H}{\Lambda}_\al(z)
+ \w{\eta\bar\eta}{\Lambda\sigma}_\al (z) \right) \Gamma = -\tfr{1}{8}
D_\al S_{\rm geom}(z)\,.
\label{Bbasis}
\end{equation}
(\ref{Bbasis}) is the basic Ward identity for a field theory coupled to the
background fields of new minimal supergravity, analogous to equation
(\ref{WIweyl}) for the old minimal formulation. In the flat space
limit $H=0$, $L_0=1$, (\ref{Bbasis}) reads
\begin{equation}
\w{\rm dyn}{\Lambda\sigma}_\al \Gamma = -\tfr{1}{2} \bar
D^\da \fdq{\Gamma}{H^{\al\da}} + \half \fdq{\Gamma}{\eta^\al}\,,
\end{equation}
which reproduces the B-type superconformal Ward identity (\ref{brconfWI2})
if we identify the B-type supercurrent $V_{\al\da}^B$ and the B breaking
$B_\al$ with
\begin{align}
V^B_{\al\da} &= 8 \fdq{\Gamma}{H^{\al\da}}\,, \label{VBdef}\\[1ex]
B_\al &= 8 \fdq{\Gamma}{\eta^\al}\,.
\label{Bdefeta}
\end{align} 
Thus we have found a curved space formalism which reduces to the B-type
Ward identity 
\begin{equation}
\w{\rm dyn}{\Lambda\sigma}_\al \Gamma_{\rm cl} = -\tfr{1}{16} \bar D^\da
V^B_{\al\da} + \tfr{1}{16} B_\al
\label{Bdef}
\end{equation}
in the flat space limit, and in which the supercurrent and
the breaking are directly coupled to external fields $H^{\al\da}$ and
  $\eta^\al$. Transformation properties of $V_{\al\da}^B$ and $B_\al$ may
  be obtained by functional differentiation of (\ref{Bbasis}) with respect
  to $H^{\al\da}$ and $\eta^\al$. The results are collected in the
  following theorem.

\begin{theorem}
In a model which is coupled to a real linear superfield $L$ in a Weyl
invariant way, the following Ward identities hold at $H=0$, $\phi=1$,
$L=1$:
\begin{align}
\w{\rm dyn}{\Lambda\sigma}_\al \Gamma &= -\tfr{1}{2} \bar
D^\da \fdq{\Gamma}{H^{\al\da}} + \half \fdq{\Gamma}{\eta^\al} \label{Bwardeta}
\\[0.2cm]
D^\al \fdq{\Gamma}{\eta^\al(z)} &= \bar D_\da \fdq{\Gamma}{\bar \eta_\da(z)}
\label{Kwardeta}
\\[0.2cm]
\W{\rm dyn}{\Lambda\sigma}(\Omega_{\rm conf}) \fdq{\Gamma}{H^{\al\da}(z)}
&= \delta \fdq{\Gamma}{H^{\al\da}(z)} 
 + \half \int \dV'
\Omega^\be(z') \fdq{^2\Gamma}{\eta^\be(z') \delta H^{\al\da}(z)} \nonumber
\\
& \quad +
 \half \int \dV' \bar \Omega_\db(z') \fdq{^2\Gamma}{\bar \eta_\db(z')
   \delta H^{\al\da}(z)}
\label{VBtransclass} 
\\[0.2cm]
\W{\rm dyn}{\Lambda\sigma}(\Omega_{\rm conf}) \fdq{\Gamma}{\eta^\al(z)}
&= \delta \fdq{\Gamma}{\eta^\al(z)} + \half \int \dV' \Omega^\be(z')
\fdq{^2\Gamma}{\eta^\be(z')\delta\eta^\al(z)} \nonumber \\
& \quad + \int\dV' \bar\Omega_\db(z') \fdq{^2\Gamma}{\bar\eta_\db(z')
  \delta\eta^\al(z)}  \label{Btransclass}\\
\intertext{with}
\delta \fdq{\Gamma}{H^{\al\da}(z)} &= \left(\Lambda
-\tfr{3}{2}(\sigma+\bar\sigma) \right)\fdq{\Gamma}{H^{\al\da}(z)} \\
\delta \fdq{\Gamma}{\eta^\al(z)} &= \left( \Lambda -\tfr{3}{2} \sigma
\right)  \fdq{\Gamma}{\eta^\al(z)}\,. \label{delta_ddeta}
\end{align}
\label{thm31}
\end{theorem}
$\delta^{(1)}\eta_\al$ does not occur in (\ref{VBtransclass}) because 
$\fdq{}{H^{\al\da}}\int \dV \Omega^\al \w{\eta}{\Lambda\sigma}_\al^{(1)}$
vanishes for $\Omega=\Omega_{\rm conf}$ as may be seen from
(\ref{deltaetaeins}). The discussion following theorem \ref{thm_Strans}
also applies to (\ref{Btransclass}), thus there is no contribution from
geometrical anomalies. (\ref{delta_ddeta}) also includes the Lorentz
transformation of the index $\alpha$.

(\ref{Kwardeta}) expresses the $K$-gauge invariance (\ref{etagauge}) and
shows that $B_\al$ as defined by (\ref{Bdefeta}) satisfies the usual constraint
(\ref{BSconstr2}) on the B breaking, 
\begin{equation}
D^\al B_\al - \bar D_\da \bar B^\da=0\,.
\end{equation}

\subsection{B-type supercurrent and energy-momentum tensor}
Energy-momentum tensor, supersymmetry current and $R$ current have been
given in terms of supercurrent components already in section
\ref{sec:SBWI}. Now we
reconsider the component currents from the point of view of new
minimal curved superspace. 

In the integrated B-type superconformal Ward identity
\[ \int \Omega^\al \w{\rm dyn}{\Lambda\sigma}_\al \Gamma + c.c. = 
 -\tfr{1}{16} \int \dV \left( \Omega^\al \bar D^\da V_{\al\da} + \bar
  \Omega_\da D^\al 
  V_\al{}^\da - \Omega^\al B_\al - \bar \Omega_\da \bar B^\da \right)\,,\]
there is no contribution from the B breaking if
\begin{equation}
\Omega^\al(z) = D^\al \omega(z)\,, \quad \omega=-\bar\omega \qquad
\text{or} \qquad \bar D^2 \Omega^\al(z) =0 \,.
\label{OmeganoB}
\end{equation}
Since we have super Poincar\'e and $R$ invariance, the corresponding flat
space transformation parameters can be written in this form, while for $K$,
$D$ and $S$ transformations this is not possible,
\begin{align}
\Omega^\al_{\rm conf} &= D^\al \omega + \bar D_\da \omega^{\al\da} + \theta^\al \bar \theta^2 \left(
  \half d + k^a x_a \right) -2\i \theta^2 \bar\theta^2  s^\al 
\label{OmegaB}\\
\intertext{with}
\omega &= \tfr{\i}{4} \, \theta\sigma_a \bar\theta \left[ c \, t^a
  -(\omega^{ab}-\omega^{ba}) x_b + d x^a - k^a x^2 + 2 k^b x_b x^a \right]
\nonumber \\
& \quad
 - \bar\theta^2 \theta^\al \left[q_\al - \sigma^a_{\al\da}\bar s^\da x_a
 \right] 
 + \theta^2 \bar \theta_\da \left[\bar q^\da + \sigma^{\al\da} s_\al x_a
 \right] \nonumber \\
& \quad
 -\tfr{\i}{2}
\theta^2 \bar\theta^2 r\,,\label{omegaB} \\
\omega^{\al\da} &= -\tfr{\i}{8} \bar\theta^2 \sigma_a^{\al\da} \left( (2-c)
  t^a -(\omega^{ab}-\omega^{ba})x_b+dx^a -k^ax^2+2k^bx_bx^a\right)
\label{omega2B}
\end{align}
where $t^a$, $q^\al$, $\omega^{ab}$, $r$, $d$, $k^a$, $s^\al$ are the
parameters of translations, susy transformations, Lorentz transformations,
R transformations, dilatations, special conformal and special supersymmetry
transformations respectively. $c$ is an arbitrary parameter which reflects
the fact that translations may be represented in any of the forms
$\Omega^\al=D^\al \omega$ or $\bar D^2\Omega^\al=0$.
In order to define currents $R_a$, $Q_{a\al}$ and $T_{ab}$ in terms of supercurrent components
only, we have to specify localized versions of translations, $R$ and
supersymmetry transformations such that (\ref{OmeganoB}) holds. Obviously,
it suffices to 
make the parameters $t_a$, $q_\al$ and $r$ in 
(\ref{omegaB}), (\ref{omega2B}) $x$-dependent and stay with (\ref{OmegaB}),
\begin{align}
\Omega^\al &= D^\al \left( \tfr{\i}{4} \theta\sigma^a
  \bar\theta  c\, t_a(x) - \bar\theta^2
\theta^\al q_\al(x) + \theta^2 \bar \theta_\da \bar q^\da(x) -\tfr{\i}{2}
\theta^2 \bar\theta^2 r(x)\right)\nonumber \\
& \quad + \bar D_\da \left( -\tfr{\i}{8} \bar \theta^2 \sigma_a^{\al\da}
  (2-c) t^a(x)\right) \,.
\label{Blocparam}
\end{align}
The Ward operators and conserved currents are then given by
\begin{align}
\half \int\dV  \left( \Omega^\al \w{\rm dyn}{\Lambda\sigma}_\al + \bar
 \Omega_\da \wb{\rm dyn}{\Lambda\sigma}^{\da} \right) 
&\equiv \int\dx \Bigl( t^a(x) \w{\rm dyn}{P}_{a}(x)\nonumber \\
&\qquad  + q^\al(x) \w{\rm
 dyn}{Q}_\al(x) + \bar q_\da(x) \w{\rm dyn}{\bar Q}^\da(x) \nonumber \\
&\qquad + r(x) \w{\rm dyn}{R}(x) 
\Bigr)\\ 
-\tfr{1}{32} \int\dV \left( \Omega^\al \bar D^\da  V_{\al\da}  - \bar
 \Omega_\da D_\al V^{\al\da} \right)
&\equiv \int \dx \Bigl(
 t^a(x) \pr^b T_{ab}
\nonumber \\
&\qquad  + q^\al(x) \pr^b Q_{b\al} + \bar q_\da(x) \pr^b \bar
 Q_b{}^\da \nonumber \\
&\qquad + r(x) \pr^b R_b(x) \Bigr) \,.
\label{Bcompcurrdef}
\end{align}
(\ref{Bcompcurrdef}) together with (\ref{Blocparam}) yields the component
currents
\begin{align}
 T_{ab}(x) &= -\tfr{1}{8}  \left( (2-c) v_{ab} +c\,
   v_{ba}(x) \right) \,,\label{nonsymmem} \\ 
Q_{a\al}(x) &= \tfr{\i}{2}\,\chi_{a\al}(x) \\
R_a(x) &= \half C_a(x)\,,
\end{align}
where $v_{ab}$, $\chi_{a\al}$ and $C_a$ are the supercurrent components as
defined in (\ref{Vcomp}). 
We may choose $c=1$ in order to get a symmetric energy momentum tensor
\[ T_{ab} = - \quar v_{(ab)}\,.\]
This reproduces the currents (\ref{RQT_Bdef}) up to
trivial factors. 

Since the energy momentum tensor is given directly by the
$\theta\bar\theta$ component of the supercurrent, the mechanism of its
coupling to the 
vierbein should be quite different from the S formulation.
In new minimal supergravity we have $J=0$, and the remaining background
fields may in a Wess-Zumino gauge be written as
\begin{align}
H^b &= \theta\sigma^a \bar\theta h_a{}^b -\i \bar\theta^2 \theta^\al
\Psi^b{}_\al + \i \theta^2\bar\theta_\da \bar\Psi^{b\da} +
\theta^2\bar\theta^2 A^b \label{WGH} \\
\eta_\al &= \e^{-\i\theta\sigma^a\bar\theta\pr_a} \theta^\be
\eta_{\al\be}(x) \qquad \text{with } \eta_{\al\be}=\eta_{\be\al}\,.
\label{WGeta}
\end{align}
The metric tensor is given in terms of the linearized vierbein $h^{ab}$ by
\begin{equation}
g_{mn} = \eta_{mn} - 2 h_{(mn)} +O(h^2)\,.
\label{gmndef}
\end{equation}
Since $h_{ab}$ occurs only in $H^b$, it clearly couples directly to the
$\theta\bar\theta$-component of $V_b$, i.e. to the non-symmetric energy
momentum tensor (\ref{nonsymmem}).
Correspondingly, $\Psi_a{}^\al$ couples to $Q^a{}_\al$ and $A_a$ to $R^a$.
Using (\ref{gmndef}), we obtain the {\em off-shell}\/ expressions for the
currents 
\begin{gather}
T_{(mn)} = -2 \left. \fdq{\Gamma}{g^{mn}} \right|_{\mbox{flat space}}  \, ,
\quad 
Q_{a\al} = 2 \left. \fdq{\Gamma}{\Psi^{a\al}} \right|_{\mbox{flat space}} 
 \, , \quad
R_a =  \left. \half \fdq{\Gamma}{A^a} \right|_{\mbox{flat space}}  \,.
\label{currentcouplingsB}
\end{gather}
Though the definition of $T_{ab}$, $Q_{a\al}$ and $R_a$ in terms of
supercurrent components is different in the S and B cases, the respective
currents always couple correctly to the component background fields. In the
S case, however, the coupling relations hold only on-shell \cite{ERS3}.

The Wess-Zumino gauge expression (\ref{WGeta}) for $\eta_\al$ translates
into a corresponding expression for $L_0$, 
\begin{align}
L_0 &= \half \left( D^\al \eta_\al + \bar D_\da \bar \eta^\da \right)
= \theta\sigma^b \bar\theta\, \pr^a B_{ab}\\
\intertext{with}
B_{ab}&= -\quar  (\sigma_{ab})_\al{}^\be
  \eta^\al{}_\be + c.c.\,.
\label{Babdef}
\end{align}
In order to clarify the role of the antisymmetric tensor field $B_{ab}$, we
consider the antisymmetric part of the $\theta$-component of the
superconformal Ward identity (\ref{Bwardeta}),
\begin{equation}
\left. (\sigma^{ab})^{\al\be} D_{(\be} w_{\al)}\Gamma
\right|_{\theta=0}+c.c. = -4 \fdq{\Gamma}{h_{[ab]}} + \half \epsilon^{abcd}
\pr_{c} \fdq{\Gamma}{A^{d}} - \fdq{\Gamma}{B_{ab}} \,.
\label{shiftidentity}
\end{equation}
This shows that on-shell, $B_{ab}$ does not couple to an
independent field. (\ref{shiftidentity}) rather expresses an invariance
with respect to a kind of shift in the external fields, similar to the {\em
  shift identity} considered in \cite{PW}. In \cite{PW}, the new minimal
supergravity fields $e_a{}^m$, $\Psi_m{}^\al$ and $B_{mn}$ are used as
background fields for a Yang-Mills theory in the component approach on tree
level. Since, however, all transformations are formulated as BRS
transformations with corresponding ghost fields, a concise comparison seems
difficult. Most likely, the two approaches -- if they are both formulated
in the BRS language and for the same model -- are equivalent in the sense
that the Slavnov-Taylor identities coincide at least modulo equations of
motion. 

$L_0$ is invariant under the K-gauge transformation (\ref{etagauge}) which
reads in the Wess-Zumino gauge 
\begin{equation}
\delta B_{ab} = \pr_b f_a - \pr_a f_b\,.
\end{equation}
This invariance gives rise to the Ward identity
\begin{equation}
\pr^a \fdq{\Gamma}{B^{ab}}(x) =0\,.
\label{BabWI}
\end{equation}
From (\ref{shiftidentity}) we deduce that on-shell this means
\begin{equation}
0=\pr^a \fdq{\Gamma}{B^{ab}} = 4 \pr^a \fdq{\Gamma}{h^{[ab]}} =4 \pr^a T_{[ab]}\,.
\end{equation}
This shows again that the antisymmetric part of the energy-momentum tensor
is separately conserved,
corresponding to the fact that the parameter $\Omega^\al$ (\ref{Blocparam})
with $c=1$
which leads to a symmetric energy momentum tensor,
consists of two parts. One part fulfills $\bar D^2 \Omega^\al=0$, and the
conservation of the corresponding part of the energy momentum tensor is due
to the chirality of $B_\al$. The second part has the form $\Omega^\al=D^\al
\omega$. Its contribution to the energy momentum tensor is conserved
because of the K gauge Ward identity (\ref{etagauge}). Thus the existence of a
conserved {\it symmetric} energy momentum tensor originates from
(\ref{etagauge}),
which in the Wess-Zumino gauge is expressed by (\ref{BabWI}).

\section{Quantized Wess-Zumino model}
\label{sec:quant}
\setcounter{equation}{0}
\setcounter{table}{0}

It has been shown in \cite{ERS1} that the massless Wess-Zumino model
can be quantized in an $R$-invariant way. Since, however, this
$R$-invariance is only manifest in the B formulation, we would like to
apply the formalism of the previous section to the Wess-Zumino model. 
The classical action for the massless Wess-Zumino model on flat space is
given by
\begin{equation}
\Gamma_{\rm cl} = \tfr{1}{16} \int\dV A \bar A + \tfr{g}{48} \int \dS A^3 +
\tfr{g}{48} \int \dSb \bar A^3\,.
\end{equation}
For quantization we use BPHZ renormalization in its generalization to
massless theories by Lowenstein and Zimmermann \cite{bphz}. Thus
we have to include an auxiliary mass term 
\begin{equation}
\tfr{1}{8} M (s-1) \int \dS A^2 + \tfr{1}{8} M(s-1) \int \dSb \bar A^2\,,
\end{equation}
which breaks superconformal invariance resp. -- on curved space -- Weyl
invariance. The parameter $s$ takes part in the BPHZ subtractions like an
external momentum. Therefore the limit $s=1$ may not be taken naively, but
we have to use a Zimmermann identity instead,
\begin{equation}
\left. [M(s-1) A^2] \cdot \Gamma \right|_{s=1} = \sum_\Delta \left. u_\Delta
[\Delta] \cdot \Gamma  \right|_{s=1} \,. \label{ZIgeneral}
\end{equation}
Here, the sum extends over all possible insertions of dimension 3 except
for the mass term iteslf. 
$\Gamma$ is the vertex functional, i.e. the generating functional for 1PI
Green functions, and the symbol $[\Delta]\cdot \Gamma$ denotes an insertion
of the composite operator $[\Delta]$ which is defined by Zimmermann's
normal product algorithm \cite{composite}.
Equation (\ref{ZIgeneral}) is the source of
superconformal anomalies. 

The most general diffeomorphism and parity invariant effective action (in
the sense of Zimmermann) on
old minimal curved superspace is given by 
\begin{align} \label{WZcq}
\Geff =\,& \tfr{1}{16}\,\hat z\, \Ikin \,-\,\tfr{1}{8}
\,\left(\Im+\bar\Im\right) \,+\, \tfr{1}{48}\,\hat g\,
\left(\Ig+\bar\Ig\right) \,+\, \tfr{1}{8}\,\hat\xi\,
\left(\Ixi+\bar\Ixi\right) \nonumber\\ &+ \,\tfr{1}{8}\,\hat\lambda_1
\, \left(\Ii+\bar\Ii\right) \,+\, \tfr{1}{8}\,\hat\lambda_2\,
\left(\Iii+\bar\Iii\right)
\end{align}
with
\begin{align}
\Ikin &= \int\dV E^{-1} A \, {\rm e}^{2iH}\bar A & \Ig &= \int\dS
\phi^3A^3 \nonumber\\ \Im &= \int\dS \phi^3\, M(s-1)A^2 & \Ixi &=
\int\dS \phi^3 RA^2 \label{omint} \\ \Ii &= \int\dS \phi^3 R^2 A & \Iii &= \int\dV
E^{-1} A \, {\rm e}^{2iH}\bar R \, .\nonumber
\end{align}
The dynamical fields $A, \bar A$ are quantized, whereas the background
fields $H,J, \bar J$ are treated as classical, i.e. non-propagating.
The counterterm coefficients $\hat z$, $\hat g$, $\hat \xi$,
$\hat\lambda_1$ and $\hat\lambda_2$ are power series in $\hbar$. $\hat z$
and $\hat g$ are fixed by the normalization conditions
\begin{gather} \label{zgcond}
\Gamma_{A \bar A} \Big|_{p^2 = - \mu^2, s=1, \theta=0} =
\tfr{1}{16} \, , \qquad \pr_{\theta_1 }{}^{\!\! 2} \pr_{\theta_2
}{}^{\!\! 2} \Gamma_{A A A} \Big|_{p^2 = q^2 = (p+q)^2 =- \mu^2,
s=1, \theta=0} = {\ts \frac{1}{8}} \, g \, ,
\end{gather}
where an index $A$ means functional differentiation with respect to $A$.
The remaining counterterm coefficients are determined by demanding $R$
invariance of the vertex functional.
This $R$ invariant theory has been
considered in detail in \cite{ERS1, ERS3}. In particular, it has been shown
that the dynamical superconformal anomalies may be parametrized by the
$\beta$ and $\gamma$ functions. However, the theory is not invariant under
$R$ transformations involving an anomalous dimension $\gamma$. Since we
would like to apply the formalism of the previous section which relies
partly on $R$ invariance, it is favourable here not to introduce an
anomalous dimension. The Weyl Ward identity then reads (using the notations
of \cite{ERS1}) 
\begin{align}
\left. \w{}{\sigma}(z) \Gamma \right|_{s=1} &= - \left. \tfr{3}{2} [S(z)] \cdot
\Gamma \right|_{s=1} - \tfr{3}{2} S_{\rm geom}(z)\,, \\
-\tfr{3}{2} S &= -\tfr{1}{8} \ukin \Lkin - \tfr{1}{16} (\uxi + \uxib) (\Lxi
  + \Lxib) \nonumber \\
& \quad - \tfr{1}{16} (\ui +\uib) (\Li +\Lib) - \tfr{1}{16} (\uii +
  \uiib) (\Lii + \Liib) - \tfr{1}{8} \ubox \Lbox\,.
\label{Sexplicit}
\end{align}
The $\L$-terms are collected in table \ref{tab_WZS_locmon}.
\mytable{ \abovedisplayskip0cm\belowdisplayskip0cm \vspace*{-0.2cm}
\begin{align}
\Lkin &=  \phi^3 A \left(\Db^2 + R \right) \e^{2\i H}\bar {A} & 
\Lg &=\phi^3 A^3  \nonumber \\
\Lm &= \phi^3 A^2  & 
\Lbox&= \phi^3\left( \Db^2+R \right) \e^{2\i H}\D^2\e^{-2\i H} \, A  \rule{0cm}{3ex} \nonumber \\
\Lxi &= \phi^3 RA^2 & 
\Lxib &= \phi^3 \left( \Db^2 + R \right)  \e^{2\i H}\bar {A}^2  \rule{0cm}{3ex} \nonumber \\
\Li &= \phi^3 R^2 A & 
\Lib &=  \phi^3 \left( \Db^2 + R \right) \e^{2\i H}(\bar A\bar R)  \rule{0cm}{3ex} \nonumber \\
\Lii &= \phi^3 \left( \Db^2 + R \right) \e^{2\i H}(\D^2+\bar R) \e^{-2\i H}A & 
\Liib &= \phi^3 R \left( \Db^2 + R \right) \e^{2\i H}\bar A  \,,\rule{0cm}{3ex} \nonumber 
\end{align}}{Local Field Monomials}{tab_WZS_locmon}
The Zimmermann coefficients $u$ are related to to the counterterm
coefficients $\hat\xi$, $\hat\lambda_1$, $\hat\lambda_2$ by
\begin{align}
\hat\xi &= \half (\uxib-\uxi)\,, \nonumber \\
\hat\lambda_1 &= \tfr{1}{4} (\uib-\ui)\,, \label{Rfix}\\
\hat\lambda_2 &= \half(\uii-\uiib)\,. \nonumber
\end{align}

We proceed by introducing an additional external field $L$ into the 
model in the way  described in section \ref{sec:weylcoupling}. For the
classical 
theory there exists only the trivial solution which has zero mass and
doesn't depend on $L$ at all. For the quantized model, however, it is a
non-trivial problem to establish the Weyl invariant coupling of $L$.

\subsection{Weyl invariant coupling of \boldmath$L$\unboldmath}
When we introduce an additional external field, the effective action
comprises all diffeomorphism and parity invariant terms of dimension 3
which can be built from the dynamical fields $A$, $\bar A$ and from $H$,
$J$ and $L$. Since $L$ is dimensionless, there are several infinite towers
of terms involving $L^n$. In order to keep the discussion digestible, we
decompose the effective action into three parts,
\begin{equation}
\Geff = \Geff^{\rm dyn}+ \Geff^{\rm lin}+ \Geff^{\rm geom}\,,
\label{WZB_Geff}
\end{equation}
where $\Geff^{\rm dyn}$ contains terms which are at least quadratic in $A$,
$\bar A$; $\Geff^{\rm lin}$ contains terms linear in $A$, $\bar A$ and
$\Geff^{\rm geom}$ comprises terms which depend only on the background
fields $H$, $J$ and $L$.
The treatment of linear terms is postponed until section
\ref{sec_WZBlin}. ***
Since we will consider only 
transformation properties of single insertions, $\Geff^{\rm geom}$ is not
relevant and will not be considered here.
 
A basis of integrated field monomials -- which is needed to  write down the
most general $\Geff^{\rm dyn}$ -- may be found in table \ref{tab_WZBt1}.
We see that the terms $\Ikin$ and $\Ixi$ from (\ref{omint}) are
extended by an additional factor $l^n$, while it is not possible to derive
$l$-dependent terms from $I_M$ and $I_g$ due to chirality. 
$l$ is defined by $l=\e^{\i H} \tl = \log L$. It should be clear, however,
that the independent fields are $A$, $\bar A$, $H^{\al\da}$, $J$, $\bar J$
and $\tl$.

The effective action is given by
\begin{equation}
\Geff^{\rm dyn} = \sum_{n=0}^\infty \left( \tfr{1}{16} \hat z^{(n)} (\nIkin{n}+\bar
  \nIkin{n} ) + \tfr{1}{8}\hat \xi^{(n)} (\nIxi{n} + \bar \nIxi{n}) \right)
  -\tfr{1}{8} (\Im+\bar \Im) + \tfr{\hat g}{48} (\Ig+\bar \Ig) \,.
\label{Geffdyn}
\end{equation}
According to the quantum action principle, the Weyl variation of the vertex
functional $\Gamma$ is given in terms of the Weyl variation of the
effective action,
\begin{equation}
\w{}{\sigma} \Gamma = \left[ \w{}{\sigma}\Geff \right] \cdot \Gamma\,.
\end{equation}
\mytable{\abovedisplayskip0cm\belowdisplayskip0cm
\begin{align}
\nIkin{n} &= \int\dV E^{-1} A \e^{2\i H}\bar A \,l^n &
\nIxi{n} &= \int \dV E^{-1} A^2 \,l^n \nonumber \\
\Im&= \int \dS \phi^3 M(s-1) A^2 
 & \Ig &= \int \dS \phi^3 A^3 \nonumber \\[2ex]
\nLkin{n} &= \phi^3 \Pc (A \e^{2\i H}\bar A \, l^n) \nonumber \\
\nLxi{n} &= \phi^3\Pc (A^2 \, l^n) & \nLxib{n} &= \phi^3 \Pc(l^n \, \e^{2\i
  H} \bar A^2 )
\nonumber \\
\Lm &= M(s-1) \phi^3 A^2 & \Lg &= \phi^3A^3 \nonumber
\end{align}}{Local and integrated field monomials of order $\ge 2$ in
  $A$, $\bar A$}{tab_WZBt1}

The Weyl variation of the effective action is
\begin{align}
\w{}{\sigma} \,\,\Geff^{\rm dyn} &= -\tfr{1}{16} \sum_{n=0}^\infty (n+1) \hat
z^{(n+1)} \nLkin{n}\,\,  -\tfr{1}{8} \Lm\nonumber \\
&\quad -\tfr{1}{8} \sum_{n=0}^\infty \left( \hat \xi^{(n)} (\nLxi{n}
  -\nLxib{n}) + (n+1) \hat\xi^{(n+1)} (\nLxi{n} + \nLxib{n}) \right)\,.
\label{WZB_weyldyn}
\end{align}

In order to evaluate $[\w{}{\sigma} \,\Geff^{\rm dyn}]\cdot\Gamma$, a
Zimmermann identity has to be used.
The Zimmermann identity (\ref{ZIgeneral}) for $[M(s-1)\phi^3 A^2]\cdot \Gamma$ contains all
possible terms of dimension $3$ which we divide into three parts,
\begin{equation}
\left. \left[ M(s-1) \phi^3 A^2 \right] \cdot \Gamma \right|_{s=1} =
\left. \left[ \Delta_{\rm dyn}\right]  \cdot \Gamma\right|_{s=1} + \Delta_{\rm lin} + \Delta_\geom \,.
\label{ZIsplit}
\end{equation}
All terms giving rise to true, non-local insertions are collected in
$\Delta_{\rm dyn}$, while $\Delta_{\rm lin}$ comprises terms linear in $A$
and $\Delta_\geom$ contains purely geometrical terms,
\begin{equation}
\Delta_{\rm dyn} = \sum_{n=0}^\infty \left\{ \nukin{n} \nLkin{n} + \nuxi{n}
  \nLxi{n} + \nuxib{n} \nLxib{n} \right\}\,.
\label{Delta_B_dyn}
\end{equation}
The Weyl breaking $S$ is also decomposed into three parts
\begin{equation}
\left. \w{}{\sigma} \, \Gamma \right|_{s=1} = -\tfr{3}{2}
\left. [S]\cdot\Gamma \right|_{s=1} = -\tfr{3}{2} \left. [S_{\rm
    dyn}]\cdot\Gamma \right|_{s=1} -\tfr{3}{2} S_{\rm lin} -\tfr{3}{2}
S_{\rm geom}\,,
\label{Sdecompose}
\end{equation}
where $-\tfr{3}{2} S_{\rm dyn}$ is given by the sum of
$-\tfr{1}{8} \Delta_{\rm dyn}$ and the terms from (\ref{WZB_weyldyn})
except for the  mass term,
\begin{align}
-\tfr{3}{2} S_{\rm dyn} &= -\tfr{1}{8} \Delta_{\rm dyn}  -\tfr{1}{16} \sum_{n=0}^\infty (n+1) \hat
z^{(n+1)} \nLkin{n} \nonumber \\
&\quad -\tfr{1}{8} \sum_{n=0}^\infty \left( \hat \xi^{(n)} (\nLxi{n}
  -\nLxib{n}) + (n+1) \hat\xi^{(n+1)} (\nLxi{n} + \nLxib{n}) \right)\,.
\label{WZB_weyldyn2}
\end{align}

As explained in section \ref{sec:weylcoupling}, we would like to have a Weyl
invariant theory (up to geometrical breaking terms), such that the
S-breaking is coupled to the external field $\tl$ according to equation
(\ref{Slcoupling}) . 
The basic idea is to choose the infinitely many counterterm coefficients in
such a way that the complete breaking $\Delta_{\rm dyn}$ is absorbed.
This Weyl invariant coupling is established by the following theorem.

\begin{theorem}
There exists a unique choice of $\hat z^{(n)}$ ($n\ge1$),
$\hat\xi^{(n)}$ ($n\ge 0$) and of the linear {counter\-term} coefficients such that
\[ \left. \w{}{\sigma}\,\,\Gamma \right|_{s=1} = -\tfr{3}{2} S_\geom(H,J,\tl)\,, \]
where $S_\geom$ is independent of $A$, $\bar A$. The only remaining free
parameters are $\hat z^{(0)}$, $\hat g$ and $M$. \\
To zeroth order in $l$, the theory coincides with the R-invariant theory
on old minimal superspace, while to first order the effective action is
given by 
\begin{align}
\left.\Geff\right|_{l^1} &=
-\tfr{1}{8} \ukin \int\dV E^{-1} l\, A \e^{2\i H} \bar A  - \tfr{1}{16} (\uxi+\uxib)
\int \dV E^{-1} l \, (A^2+ \e^{2\i H} \bar A^2)
\nonumber \\
& \quad  -\tfr{1}{16} (\ui+\uib) \int\dV E^{-1} l\,
\left(RA+ \e^{2\i H}(\bar R \bar A) \right)
\nonumber \\
&\quad +\tfr{1}{16} (\uii+\uiib) \int \dV E^{-1} l\, \left( \e^{2\i H} \Pac
  \e^{-2\i H} A + \Pc \e^{2\i H} \bar A \right)
\nonumber \\
& \quad +\tfr{1}{8} \ubox \int \dV E^{-1} l \,( \e^{2\i H}\D^2 \e^{-2\i H} A + \Db^2
\e^{2\i H} \bar A) \,.
\nonumber 
\end{align}
\label{thm61}
\end{theorem}
\begin{proof}
In order to reveal hidden dependencies between the Zimmermann coefficients,
  we make use of the commutation relation
  $[\wb{}{\sigma}(z')\,,\w{}{\sigma}(z)]=0$ which implies that 
\begin{equation}
\wb{}{\sigma}(z')\, S_{\rm dyn}(z)
= \w{}{\sigma}(z)\, \bar S_{\rm dyn}(z') + O(\hbar^k)\,,
\label{WZB_consistency_xi}
\end{equation}
where $k-1$ is the lowest non-vanishing $\hbar$-order of $S_{\rm
  dyn}$,
\begin{equation}
S_{\rm dyn}= O(\hbar^{k-1})\,.
\end{equation}
(\ref{WZB_consistency_xi}) is equivalent to the consistency relations
\begin{equation}
\nuxi{n} - (n+1) \nuxi{n+1} = -\nuxib{n} - (n+1) \nuxib{n+1} +O(\hbar^k)
\qquad (n\ge 0) \,.
\label{WZB_xiconsistency}
\end{equation}
Note that the superscripts $^{(n)}$ denote the order in $l$, not in $\hbar$.
In order to solve equation (\ref{WZB_xiconsistency}) we define
\begin{align}
\vxi{0} &\equiv \half (\nuxi{0} -\nuxib{0}) \label{vxinull}\\
\vxi{n+1} &\equiv \tfr{1}{n+1} (\nuxi{n} -\vxi{n})\,, \label{vxin}
\end{align}
or inverted
\begin{equation}
\nuxib{0}=\vxi{1}-\vxi{0} \qquad \nuxi{n}=\vxi{n}+(n+1)\vxi{n+1} \quad
(n\ge 0)\,.
\end{equation}
Equation (\ref{WZB_consistency_xi}) determines all $\nuxib{n}$ in terms of
the $\vxi{n}$, 
\begin{equation}
\nuxib{n}= -\vxi{n} + (n+1) \vxi{n+1} + O(\hbar^k)\,,
\end{equation}
such that $\Delta_{\rm dyn}$ may be rewritten as
\begin{equation}
\Delta_{\rm dyn} = \sum_{n=0}^\infty \left(
\nukin{n} \nIkin{n} + \vxi{n} (\nLxi{n}-\nLxib{n}) + (n+1) \vxi{n+1}
(\nLxi{n}+\nLxib{n}) \right) + O(\hbar^k)
\end{equation} 
Comparison with (\ref{WZB_weyldyn2}) shows that all breaking terms are Weyl
variations and can be absorbed into $\Geff^{\rm dyn}$ by choosing
\begin{equation}
\hat z^{(n)} = -\tfr{2}{n} \nukin{n-1} \quad (n\ge 1)\,, \qquad \hat\xi^{(n)}
= -\vxi{n} \quad (n\ge 0)\,. \label{nzxiabsorb}
\end{equation}
Thus we have
\begin{equation}
\left. \w{}{\sigma}\,\,\Gamma \right|_{s=1} = -\tfr{3}{2} S_\geom(H,J,\tl)+
\text{terms linear in $A$, $\bar A$} + O(\hbar^k)\,,
\end{equation}
\vspace*{-0.5ex}i.e.\vspace*{-0.5ex}
\begin{equation}
S_{\rm dyn}= O(\hbar^k)\,.
\end{equation}
Now the lowest non-vanishing order of $S_{\rm dyn}$ is no longer
$k-1$ but $k$, and in this way $k$ can be pushed higher and higher until
$S_{\rm dyn}$=0 to all orders in $\hbar$.

Putting $l=0$, the Zimmermann identity (\ref{ZIsplit}) becomes identical to
the Zimmermann identity on old minimal curved superspace
\cite{ERS1}. Comparison of coefficients yields 
\[ \ukin=\nukin{0}\,,\qquad \uxi=\nuxi{0}\,,\qquad \uxib=\nuxib{0}\,.\]
Correspondingly, we have $\hat z=\hat z^{(0)}$, $\hat\xi=\hat\xi^{(0)}$.
(\ref{nzxiabsorb}) together with (\ref{vxinull}) yields the same value for $\hat \xi$ which
is also obtained by requiring R invariance. Thus to zeroth order in $l$,
the model is equivalent to the R invariant theory of \cite{ERS1}.
The coefficients of the first order terms in $l$ are obtained from
(\ref{vxin}), (\ref{nzxiabsorb}) as
\[
\hat\xi^{(1)} = -\half (\uxi+\uxib)\,, \qquad \hat z^{(1)} = -2\ukin\,.
\]

The linear terms will be treated separately in section \ref{sec_WZBlin}.
\\
\end{proof}

Clearly, $\hat z^{(0)}$ and $\hat g$ are the same as $\hat z$ and $\hat g$
in the theory without $L$ and are fixed by the usual normalization
conditions (\ref{zgcond}).

\subsection[Transformation of $S$]{Transformation of \boldmath$S$\unboldmath}
Having achieved the Weyl invariant coupling of $L$, (\ref{Stransclass}) may
now be applied to the vertex functional of the Wess-Zumino model. 
Using the expression for $\Geff|_{l^1}$ of theorem \ref{thm61}, it may be
checked explicitly that the $S$ breaking (\ref{Sexplicit}) is coupled to
$\tl$ according to (\ref{Slcoupling}). 
On flat space, (\ref{Sexplicit}) reduces to
\begin{equation}
-\tfr{3}{2} S = -\tfr{1}{8} \ukin \bar D^2 (A \bar A) - \tfr{1}{16} (\uxi+\uxib) \bar
D^2 \bar A^2 + (\uii+\uiib+2\ubox) \Box A\,.
\label{Sflat}
\end{equation}
(\ref{Stransclass}) yields the transformation properties of the $S$
breaking (\ref{Sflat}) for the flat space theory at $s=1$,
\begin{align}
\W{A}{}(\Omega_{\rm conf}) \, [S(z)]\cdot \Gamma &= \left[
  (\Lambda_c-3\sigma) S(z) \right]\cdot \Gamma  
 -\tfr{3}{2} \int\dS' \sigma(z') \left\{ S(z')\cdot S(z) \right\} \cdot
  \Gamma
\nonumber \\
&\quad -\tfr{3}{2} \int\dSb' \bar \sigma(z') \left\{ \bar S(z') \cdot S(z)
  \right\} \cdot \Gamma\,.
\label{Stransquant}
\end{align}
The double insertion $\{S(z_1) \cdot S(z_2)\}$ is defined in the spirit of
\cite{ERS3} by
\begin{align}
\{S(z_1) \cdot S(z_2)\} \cdot\Gamma & = \left(-\tfr{2}{3}\right)^2 \bar D_1^2
\bar D_2^2 \fdq{^2\Gamma}{\tl(z_1) \delta \tl(z_2)} \nonumber \\
&= [S(z_1)]\cdot [S(z_2)]\cdot \Gamma + \left[  \left(-\tfr{2}{3}\right)^2
  \bar D_1^2 \bar D_2^2  \fdq{^2\Geff}{\tl(z_1) \delta  \tl(z_2)}
\right]\cdot \Gamma\,, \nonumber
\end{align}
which implies that the second derivative of $\Geff$ with respect to $\tl$
is involved. Terms of second order in $\tl$ in $\Geff$ contain the Zimmermann
coefficients $\nukin{1}$, $\nuxi{1}$, \dots which are not present in flat
space nor in the old minimal curved space formalism of \cite{ERS1}. Actual
values  
for these Zimmermann coefficients can be obtained by calculating Feynman
diagrams with one external $\tl$-leg. 

(\ref{Stransquant}) holds for any choice of geometrical counter\-terms
$\Geff^{\rm geom}$. The geometrical counterterms of first order in $\tl$
contribute to $S$ itself, while the second order terms contribute to the
double insertion. Due to diffeomorphism invariance, the first and second
order terms are related in exactly such a way that (\ref{Stransquant}) is
always true. 

We pass over from $\Gamma$ to $Z$ and obtain the transformation properties
of Green functions.
\begin{theorem}
The superconformal transformation properties for the $S$ breaking
(\ref{Sflat}) of the
massless flat-space Wess-Zumino model are given by
\begin{align}
\delta \l \T S(z) \, X\r &= \tfr{3}{2}\i \int\dS\!' \sigma(z') \, \l \T
\{ S(z') \cdot S(z)\} \, X \r \nonumber \\
& \quad + \tfr{3}{2} \i \int \dSb\!' \bar \sigma(z')
\, \l \T \{ \bar S(z') \cdot S(z) \} \, X \r  \,,
\label{StransGreen} \\
\intertext{where}
X &\equiv A(z_1) \dots A(z_n) \bar A(z_1') \dots \bar A(z_m')\,
\end{align}
is an arbitrary number of elementary fields. $\l \T \dots \r$ is the vacuum
expectation value of the time ordered product.
Here the superconformal transformation $\delta$ of the Green functions is
defined by
\begin{align}
\delta \, \l \T S(z) \, X \r & \equiv \l \T \delta S(z) \,
X \r \nonumber \\
&\quad  + \sum_{k=1}^n \l \T S(z) \, A(z_1) \dots \left( \delta A(z_k)\right) \dots
  A(z_n) \, \bar A(z_1') \dots \bar A(z_m')\r \nonumber \\
& \quad + \sum_{k=1}^m \l \T S(z) \, A(z_1) \dots
  A(z_n) \, \bar A(z_1') \dots \left( \delta \bar A(z_k')\right) \dots \bar
  A(z_m')\r \,, 
 \\[-0.6cm]
\intertext{with \vspace{-0.4cm}}
\delta S &= \left( \Lambda_c -3\sigma \right) S \,, \nonumber \\
\delta A &= \left( \Lambda_c - \sigma \right) A \,, \nonumber \\
\delta \bar A &=  \left( \bar \Lambda_c - \bar \sigma \right) \bar A
\,. \nonumber
\end{align}
\label{thm_Stransquant}
\end{theorem}
Since we have not assigned any 
anomalous dimension to $A$, we have -- in addition to super Poincar\'e
invariance -- exact $R$-invariance, i.e.
\begin{equation}
\delta^{P,Q,M,R} \l \T S(z)\, X \r =0\,.
\end{equation}

\subsection{B formulation}
In order to obtain the B formulation, $L$ has to be restricted to be a
linear field. This means, however, that the basis of local field monomials
presented in tables \ref{tab_WZBt1}, \ref{tab_WZBt2a}, \ref{tab_WZBt2b} is
no longer linearly independent. It turns out that $\nLvi{n}$, $\nLvib{n}$,
$\nLvii{n}$, $\nLviib{n}$ may be expressed in terms of the remaining basis
elements. In the Zimmermann terms $\Delta_{\rm I}^{\rm lin}$ (\ref{DeltalinI}) and
  $\Delta_{\rm II}^{\rm lin}$ (\ref{DeltalinII}), these terms have to be
  omitted. $\Delta_{\rm dyn}$ remains 
  unchanged. Correspondingly, the most general effective action is given by
\[ \Geff = \Geff^{\rm dyn} + \Geff^{\rm lin \,\, I} + \Geff^{\rm lin \,\,
  II}\]
with $\Geff^{\rm dyn}$ as in (\ref{Geffdyn}) and $\Geff^{\rm lin}$ as in
(\ref{GefflinI}), (\ref{GefflinII}) with $\lambda_{\rm
  Ib}^{(n)}=\lambda_{\rm IIc}^{(n)}=0$.
The consistency conditions for the new Zimmermann identities are the same
as in (\ref{consistencyI}) and (\ref{consistencyII}), but with
$\nuvi{n}=\nuvib{n}=\nuvii{n}=\nuviib{n}=0$. It is easy to see that still
all breaking terms can be absorbed, i.e. theorem \ref{thm61} is also valid for
linear $L$.

B-type supercurrent and B breaking may be easily calculated by using
\begin{align}
\left. \tl \right|_{L_0=1} &= -\tfr{1}{3} [D_\al, \bar D_\da] H^{\al\da} +
O(H^2) \\[0.3cm]
\left. \tl \right|_{H=0} &= \log \left( \half D^\al \eta_\al + \half \bar D_\da
  \bar \eta^\da \right)\,.
\end{align}
From (\ref{VBdef}), (\ref{Bdef}) and theorem \ref{thm61}
 it follows that 
\begin{align}
V^B_{\al\da} &= V^S_{\al\da} + \tfr{1}{24} \ukin [D_\al, \bar D_\da] (A\bar
A) - \tfr{1}{48} (\uxi+\uxib) [D_\al, \bar D_\da] (A^2+\bar A^2) \nonumber
\\
& \quad + \tfr{1}{48} (\uii+\uiib) [D_\al, \bar D_\da] (D^2 A + \bar D^2
\bar A)\,, \\[0.3cm] 
B_\al &= -\tfr{1}{2} \ukin \bar D^2 D_\al (A\bar A) + \tfr{1}{4}
(\uxi+\uxib)\bar D^2 D_\al (A^2+\bar A^2)\,,
\end{align}
where $V_{\al\da}^S$ is given by 
\begin{align}
V^S_{\al\da} &= 
-\tfr{1}{6} \hat z \left( D_\al A \bar D_\da \bar A - A D_\al \bar
  D_\da \bar A + \bar A \bar D_\da D_\al A \right) \nonumber \\
& \quad
-\tfr{1}{3} \hat\xi \{ D_\al, \bar D_\da\}\left( A^2 - \bar A^2 \right) +
\tfr{1}{3} \hat \lambda_2 \{ D_\al, \bar D_\da \} \left( D^2 A - \bar D^2
  \bar A \right) \,.
\end{align}
The conformal transformation properties of the B-type supercurrent and of
$B_\al$ are
obtained from theorem \ref{thm31}.
\begin{align}
\W{A \bar A}{\Lambda\sigma}(\Omega_{\rm conf}) [V^B_{\al\da}(z)]\cdot \Gamma =
[\delta V^B_{\al\da}(z)] \cdot\Gamma &+ \tfr{1}{16} \int\dV\!'\,
\Omega^\be(z')\, \{ B_\be(z') \cdot V^B_{\al\da}(z) \} \cdot
\Gamma\nonumber \\ 
&+ \tfr{1}{16} \int\dV\!'\,
\bar \Omega_\db(z') \,\{ \bar B^\db(z') \cdot V^B_{\al\da}(z)\} \cdot \Gamma\,,
\label{VBquanttrans_Gamma} \\[0.3cm]
\W{A \bar A}{\Lambda\sigma}(\Omega_{\rm conf}) [B_\al(z)] \cdot \Gamma
= [\delta B_\al(z)]\cdot \Gamma 
&+ \tfr{1}{16} \int \dV\!'\, \Omega^\be(z') \, \{ B_\be(z') \cdot B_\al(z)\}
\cdot \Gamma \nonumber \\
& + \tfr{1}{16} \int \dV\!'\, \bar \Omega_\db(z') \, \{ \bar B^\db(z') \cdot
B_\al(z) \} \cdot \Gamma\,,
\label{Bquanttrans_Gamma}
\end{align}
We will not calculate
transformation properties of multiple insertions of the B-type supercurrent
here. However, it is clear that for every additional insertion, higher
Zimmermann coefficients of the infinite towers in (\ref{Delta_B_dyn}),
(\ref{DeltalinI}) and (\ref{DeltalinII}) will contribute. In this sense,
multiple insertions of the B-type current acquire new anomalies for each
insertion. The same is true for multiple insertions of the B-breaking
$B_\al$. 

\begin{theorem}
The superconformal transformation properties of the B-type supercurrent
and of the breaking term $B_\al$ of the massless flat-space Wess-Zumino
model are given by
\begin{align}
\delta \l \T V^B_{\al\da}(z) \, X \r &= -\tfr{\i}{16}  \int\dV\!'\,
\Omega^\be(z')\, \l \T \{ B_\be(z') \cdot V^B_{\al\da}(z) \}\, X \r 
\nonumber \\ 
&\quad - \tfr{\i}{16} \int\dV\!'\,
\bar \Omega_\db(z') \, \l\T \{ \bar B^\db(z') \cdot V^B_{\al\da}(z)\}\, X \r 
\label{VBquanttrans} \\[0.3cm]
\delta \l \T B_\al(z)\, X\r &= 
- \tfr{\i}{16} \int \dV\!'\, \Omega^\be(z') \,\l \T \{ B_\be(z') \cdot
B_\al(z)\}\, X \r \nonumber \\
& \quad - \tfr{\i}{16} \int \dV\!'\, \bar \Omega_\db(z') \, \l \T \{ \bar
B^\db(z') \cdot B_\al(z) \}\, X\r\,.
\label{Bquanttrans}
\end{align}
$\delta V_{\al\da}^B$ and $\delta B_\al$ are given by
\begin{align}
\delta V_{\al\da}^B &= \left(\Lambda
-\tfr{3}{2}(\sigma+\bar\sigma) \right) V_{\al\da}^B \\
\delta B_\al &= \left( \Lambda -\tfr{3}{2} \sigma
\right)  B_\al\,,
\end{align}
$\delta A$, $\delta \bar A$ and $X$ are the same as in theorem \ref{thm_Stransquant}.
\label{thm_Btransquant}
\end{theorem}
Due to the properties (\ref{BSconstr}) of the breaking term $B_\al$, we
have super Poincar\'e- and $R$-invariance,
\begin{equation}
\delta ^{P,Q,M,R} \l \T V^B_{\al\da}(z) \, X \r = \delta ^{P,Q,M,R} \l \T
B_\al(z) \, X \r =0\,.
\end{equation}
The dilatational Ward identity is given by
\begin{equation}
\W{A \bar A}{D} \,[V^B_{\al\da}] \cdot \Gamma = [\delta^D V^B_{\al\da}]\cdot
\Gamma - \tfr{1}{4} \int \dx \left\{ \left. (D^\be B_\be + \bar D_\db \bar
    B^\db ) \right|_{\theta=0}
  \cdot V_{\al\da}^B \right\} \cdot \Gamma\,,
\end{equation}
and similar for the $[B_\al]$-insertion.

\section{Conclusion}
\setcounter{equation}{0}
\setcounter{table}{0}

The purpose of this paper was twofold. First, it was intended to set up a
general framework which gives control on the flat-space 
supercurrent and on the superconformal anomalies by coupling the matter
theory to a background of new minimal supergravity fields. 
Second, this formalism should be applied to the massless Wess-Zumino
model.

In section \ref{sec:weylcoupling}, the S breaking has been coupled to a real
external field $L$, and thus access to the transformation properties of $S$ was
gained. By restricting the field $L$ to be linear, one performs the
transition from old to new minimal supergravity. The choice of independent
fields in new minimal supergravity is non-trivial, since it should be
possible to vary them independently and also to have well defined
functional derivative operators, which is both not the case for the
original field $L$. However, the problem has been solved by introducing
the flat space chiral spinor field $\eta_\al$. Furthermore it turned out
that this formulation yields the B-type superconformal Ward identity in the
flat space limit, that the B-type supercurrent is directly coupled to
$H^{\al\da}$ and that the B breaking $B_\al$ is directly coupled to
$\eta^\al$.  Insertions of the B-type supercurrent as well as of the
breaking term $B_\al$ may thus be generated by functional differentiation
with respect to $H^{\al\da}$ and $\eta^\al$.

This general framework has been applied to the massless Wess-Zumino model
in section \ref{sec:quant}. The classical theory is
conformally invariant, yet acquires anomalies upon quantization due to the
necessity of introducing an auxiliary mass term. 
Renormalization of the model with the additional external field $L$
involves several infinite towers of counterterms because $L$ is
dimensionless. However, a systematical treatment of dynamical anomalies was
possible by using consistency conditions. 
Especially the linear terms require quite some work here.
Though these linear terms are of little physical importance for the flat
space theory, they might very well become important in a theory with
propagating supergravity fields. 
Furthermore their study is justified as further
evidence for the vigor of the method of Weyl invariant coupling.

As a result, all dynamical
anomalies could be absorbed into the effective action, such that the
formalism of section \ref{sec:weylcoupling} could be used to determine the
transformation properties of the $S$ breaking.  Conversely
to the case of supercurrent insertions, the transformation of insertions of
$S$ involves new anomalies originating in the infinite towers of couterterms.
In the same way the transformation of the B-type supercurrent and of
the breaking term $B_\al$ involve new anomalies.

As opposed to the S formulation of \cite{ERS1}, the B
approach is technically more 
involved but perhaps also more systematic. Despite the fact that in the B
formulation $R$ invariance is manifest, the conformal transformation
properties of the S-type supercurrent are better behaved in the sense that
all anomalies 
even of multiple insertions may be expressed in terms of finitely many
Zimmermann coefficients. Multiple insertions of the B-type supercurrent
acquire new anomalies for each additional insertion.
The clarification of these different characteristics of S- and B-type
breaking constitute one of the main aims of the paper, particularly in
view of the suitable supergravity variables.

\subsection*{Acknowledgements}
I am grateful to Klaus Sibold for many useful discussions as well as to
Wolfdieter Lang for a discussion on new minimal supergravity.

\newpage
\begin{appendix}

\renewcommand{\theequation}{\thesubsection.\arabic{equation}}

\section{Appendix}

\subsection{Infinitesimal Transformations \label{sec:infinitesimal}}
\setcounter{equation}{0}
Here we list the infinitesimal diffeomorphism and Weyl transformations for
the old minimal supergravity prepotentials $H^{\al\da}$ and $J$ as well as
for a chiral matter field $A$.

\noindent{\bf Diffeomorphisms}
\begin{align}
 \delta_\Lambda A &=  {\ts \frac{1}{4}} \bar D^2 \left(  \Omega^\alpha
 D_\alpha A  \right)
 \\[1ex] 
 \delta_\Lambda H^{\alpha\da} &= \half \bar D^\da \Omega^\alpha
 \nonumber\\ &\quad + \quar \bar D^{\dot\beta} \Omega^\beta
 \{D_\beta,\bar D_{\dot\beta} \} H^{\alpha\da} - \quar
 H^{\beta\dot\beta} \{D_\beta, \bar D_\db \} \bar D^\da \Omega^\alpha
 + \quar \bar D^2 \Omega^\beta D_\beta H^{\alpha\da}
                         \label{delta_Omega_H} \nonumber \\
      &\quad - {\ts \frac{1}{8}} H^{\gamma\dot\gamma} \{D_\gamma,\bar
D_{\dot\gamma}\} \bar D^2 \Omega^\be D_\be H^{\al\da} + {\ts
\frac{1}{24}} H^{\gamma\dot\gamma}\{D_\gamma,\bar D_{\dot\gamma}\}
\left( H^{\be\db} \{D_\be,\bar D_\db\} \bar D^\da \Omega^\al \right)
\nonumber \\ &\quad - {\ts \frac{1}{12}}
H^{\gamma\dot\gamma}\{D_\gamma,\bar D_{\dot\gamma}\} \left( \bar
D^\db\Omega^\be \{D_\beta,\bar D_{\dot\beta} \} H^{\al\da} \right)
\nonumber \\ &\quad +{ \ts \frac{1}{24}} \bar D^\db \Omega^\beta
\{D_\beta,\bar D_{\dot\beta} \} \left( H^{\gamma\dot\gamma}
\{D_\gamma,\bar D_{\dot\gamma}\} H^{\alpha\da} \right) \\ 
&\quad + c.c. \,  + O(H^3)  \nonumber \\[1ex]
\delta_\Lambda J &= \quar \bar D^2 \left( \Omega^\al D_\al J \right) +
\tfr{1}{12} \bar D^2 D^\al \Omega_\al 
\end{align}

\noindent{\bf Weyl Transformations}
\begin{align}
\delta_\sigma A &= -\sigma A \\[1ex]
\delta_\sigma H^{\al\da} &= 0 \\[1ex]
\delta_\sigma J &= \sigma  
\end{align}

\subsection{Linear terms}
\label{sec_WZBlin}
\setcounter{equation}{0}
\setcounter{table}{0}

We come back to the linear terms of equations (\ref{WZB_Geff}) and
(\ref{ZIsplit}) which have not been considered so far.
First we have to find a basis of local chiral terms of dimension 3 which are
linear  in $A$, $\bar A$. Since a chiral projection
operator $\Pc$ has to be included, there is room for two additional
spinorial derivatives in order to have dimension 3. The problem of finding
all possible placements of these two derivatives such that linearly
independent terms are obtained, is easily solved when the following
identities are taken into account,
\begin{align}
\D^\al l^m \D_\al l^n &= \frac{mn}{(m+n)(m+n-1)} \D^2 l^{m+n} -
\frac{1}{m+n-1} l^{m+n-1} \D^2 l\,, 
\label{basisreduct1} \\[0.2cm]
l^m \D^2 l^n &= \frac{n(n-1)}{(m+n)(m+n-1)}\D^2 l^{m+n} +
\frac{mn}{m+n-1} l^{m+n-1} \D^2 l\,,\label{basisreduct2}\\[0.2cm]
l \D^2 \left(Al^n\right) &= \frac{n}{n+1} \D^2\left( A l^{n+1}\right)
  + \frac{1}{n+1} \D^2 \! A \, l^{n+1} - \frac{1}{n+1} A \D^2 l^{n+1} + A
  l^n \D^2 l\,,\label{basisreduct3}
\end{align}
and similarly for $\Db^2$. In (\ref{basisreduct1}) -- (\ref{basisreduct3}),
representation changing factors $\e^{2\i H}$ and $\e^{-2\i H}$ have been
  omitted for better readability. It is a trivial task to insert them at
  the correct positions.

A possible choice of basis terms $\L$ -- separated into two
parts -- is shown in tables \ref{tab_WZBt2a} and \ref{tab_WZBt2b}. Again,
representation changing factors have been omitted. 
It is important to note that
\begin{gather}
\nLv{0} = \nLii{0} \qquad \nLvb{0}=\nLiic{0} \qquad \nLvc{0}=\nLiib{0}
\nonumber \\
\nLvi{0} =\nLvb{1} \qquad \nLvib{0}=\nLvc{1} \nonumber \\
\nLic{0}=\nLib{0}\qquad \nLvii{0} = \nLi{1} \qquad \nLviib{0}=\nLic{1}\,,
\nonumber 
\end{gather}
thus the basis is given by
\begin{align}
\nLi{n}\,, \nLib{n}\,, \nLii{n}\,,\nLiib{n}\,,\nLiic{n} & \qquad(n \ge 0)
\nonumber \\
 \nLic{n}\,,
\nLvii{n}\,, \nLviib{n}\,, \nLv{n}\,, \nLvb{n}\,, \nLvc{n}\,, \nLvi{n}\,,
\nLvib{n} &  
\qquad (n \ge 1) \,.\nonumber
\end{align}

A basis of integrated terms is also given in tables \ref{tab_WZBt2a} and
\ref{tab_WZBt2b}. 

The linear terms of the effective action $\Geff^{\rm lin}$, Zimmermann
identity $\Delta_{\rm lin}$ resp.\ Weyl
breaking $S_{\rm lin}$ may be divided in two parts which can be treated
separately,
\begin{equation}
\Geff^{\rm lin} = \Geff^{\rm lin \,\,I}+ \Geff^{\rm lin \,\,II}\,, \qquad
\Delta_{\rm lin} = \Delta_{\rm lin}^{\rm I} +  \Delta_{\rm lin}^{\rm II}\,, 
\qquad 
S_{\rm lin} = S_{\rm lin}^{\rm I} +  S_{\rm lin}^{\rm II}\,.
\end{equation}
With both parts we proceed now in complete analogy to the $\nLxi{n}$,
$\nLxib{n}$ terms in the proof of theorem \ref{thm61}.

\mytable{\abovedisplayskip0cm\belowdisplayskip0cm
\begin{align}
\nIi{n} &= \int\dV E^{-1} RA\, l^n &
\nIvii{n} &= \int\dV E^{-1} A\, \Pc l \, l^n
\nonumber \\[1ex]
\nLi{n} &= \phi^3 \Pc (RA \,l^n) &
\nLvii{n} &= \phi^3 \Pc (A \,\Pc l \, l^n) 
\nonumber \\
\nLib{n} &= \phi^3 \Pc \left(   \bar R \bar A \, l^n \right) & 
\nLviib{n} &= \phi^3 \Pc \left( \bar A \,  \Pac 
  l  \, l^n \right)
 \nonumber \\
\nLic{n} &= \phi^3 \Pac (\bar A \, l^n) &
\nonumber
\end{align}}{
Local and integrated field monomials linear in
  $A$, $\bar A$: part I}{tab_WZBt2a}

\subsection*{Part I}

For the first part we have
\begin{align}
\Geff^{\rm lin \,\, I} &= \tfr{1}{8}\sum_{n=0}^\infty \lambda_{\rm Ia}^{(n)}
  (\nIi{n}+\bar \nIi{n}) + \tfr{1}{8}\sum_{n=1}^\infty \lambda_{\rm Ib}^{(n)} (\nIvii{n} +
  \bar \nIvii{n}) 
\label{GefflinI}\\
\Delta_{\rm lin}^{\rm I} &= \sum_{n=0}^\infty \left\{ \nui{n} \nLi{n} +
  \nuib{n} \nLib{n} \right\} + \sum_{n=1}^\infty \left\{ \nuic{n} \nLic{n} +
  \nuvii{n} \nLvii{n} + \nuviib{n} \nLviib{n} \right\} \,.
\label{DeltalinI}
\end{align}

Again we evaluate the commutation relations
$[\wb{}{\sigma}(z'),\w{}{\sigma}(z)]=0$,
$[\w{}{\sigma}(z'),\w{}{\sigma}(z)]=0$.
Since linear terms are always local, 
\begin{equation}
 \wb{}{\sigma}(z')\, \Delta_{\rm lin}^{\rm
  I}(z) = \w{}{\sigma}(z)\, \bar \Delta_{\rm lin}^{\rm I}(z') \,,
\qquad  
\w{}{\sigma}(z') \Delta_{\rm lin}^{\rm
  I}(z) = \w{}{\sigma}(z) \Delta_{\rm lin}^{\rm I}(z')
\end{equation}
hold to all orders in $\hbar$.
This yields the following system of equations (which is analogous to
(\ref{WZB_xiconsistency})): 
\begin{subequations}
\label{consistencyI}
\begin{align}
\nui{n} -(n\!+\!1) \nui{n+1} + 2 \nuib{n} + (n\!+\!1) \nuib{n+1} + \nuviib{n} -
\tfr{n-2}{n} \nuvii{n-1} &\,= 0 & (n\ge2) \\[0.1cm]
\nui{n} + 2 \nuic{n} + (n\!+\!1) \nuic{n+1} + \tfr{n-2}{n} \nuvii{n-1} - \nuvii{n}
&\,=0 & (n\ge 2) \\[0.1cm]
2 \nuviib{n} + (n\!+\!1) \nuviib{n+1} + 2\nuvii{n} -(n\!+\!1) \nuvii{n+1} &\,=0 &
(n\ge 1)  \\[0.1cm]
\tfr{n-2}{n} \nuviib{n-1} -\nuviib{n} + \nuib{n} -\nuic{n} + (n\!+\!1) \nuic{n+1} &\,=0
& (n\ge 2) \\[0.1cm]
2 \nui{0} - \nui{1} +2\nuib{0} + \nuib{1} + \nuic{1} &\,=0  \\[0.1cm]
\nui{1} -2\nui{2} + 2\nuib{1} +2\nuib{2} +\nuviib{1} &\,=0  \\[0.1cm]
\nui{1} + 2\nuic{1} +2\nuic{2} + \nuviib{1} -2\nuvii{1} &\,=0\\[0.1cm]
-\nuviib{1} + \nuib{1} -\nuic{1} +2\nuic{2} &\,=0
\end{align}
\end{subequations}
In order to solve these equations we define -- similarly to
(\ref{vxinull}), (\ref{vxin}) -- the linear combinations
\begin{subequations}
\label{vdefI}
\begin{align}
\qquad \qquad\qquad \qquad\qquad \qquad\qquad \vvii{n} &\,\equiv\, \tfr{1}{4} (\nuviib{n} -\nuvii{n}) &(n \ge 1) \\[0.1cm]
\vi{0} &\,\equiv \,\tfr{1}{4} (\nuib{0}-\nui{0})  &\\[0.1cm]
\vi{1} &\,\equiv -\tfr{1}{2} (\nuib{0} + \nui{0} ) &  \\[0.1cm]
\vi{n} &\,\equiv\, \nuic{n} -\tfr{n-2}{n} \, \vvii{n-1} + \vvii{n} \qquad
&(n\ge 2) \,.
\end{align}
\end{subequations}
The above consistency conditions (\ref{consistencyI}) may then be used to
express all Zimmermann 
coefficients $u$ in terms of $\vi{n}$, $\vvii{n}$. 
The resulting breaking term $\Delta_{\rm lin}^{\rm I}$ is a Weyl variation
and may be cancelled by fixing the counterterm 
coefficients 
\begin{equation}
\lambda_{\rm Ia}^{(n)} = \vi{n}\,, \qquad \lambda_{\rm
  Ib}^{(n)}=\vvii{n}\,.
\label{linabsorbI}
\end{equation}

\mytable{\abovedisplayskip0cm\belowdisplayskip0cm
\begin{align}
\nIii{n} &= \int\dV E^{-1} \bar R A l^n
&
\nIv{n} &= \int\dV E^{-1} \Pac A \, l^n 
\nonumber \\
\nIvi{n} &= \int \dV E^{-1} A \Pac l \, l^n &
\nonumber \\[1ex]
\nLii{n} &= \phi^3 \Pc \Pac (A l^n) 
& \nLv{n} &= \phi^3 \Pc ( \Pac A l^n )
\nonumber \\
\nLiib{n} &= \phi^3 \Pc (R \bar A l^n ) 
& \nLvb{n} &= \phi^3 \Pc (A \Pac l^n )
 \nonumber \\
\nLiic{n} &= \phi^3 \Pc ( \bar R A l^n) 
& \nLvc{n} &= \phi^3 \Pc ( \Pc \bar A l^n )
\nonumber \\
\nLvi{n} &= \phi^3 \Pc ( A \Pac l \, l^n) 
& \nLvib{n} &= \phi^3 \Pc (\bar A \Pc l \, l^n ) &
\nonumber
\end{align}}{
Local and integrated field monomials linear in
  $A$, $\bar A$: part II}{tab_WZBt2b}

\subsection*{Part II}

Part II is treated in exactly the same way as part I. Effective action and
Zimmermann terms are given by
\begin{align}
\Geff^{\rm lin \,\, II} &= \tfr{1}{8}\sum_{n=0}^\infty \lambda_{\rm
  IIa}^{(n)} (\nIii{n} + \bar \nIii{n}) + \tfr{1}{8}\sum_{n=1}^\infty
  \lambda_{\rm  IIb}^{(n)} (\nIv{n} + \bar \nIv{n}) 
 + \tfr{1}{8}\sum_{n=1}^\infty
  \lambda_{\rm IIc}^{(n)} (\nIvi{n} + \bar \nIvi{n}) 
\label{GefflinII}
\\
\Delta_{\rm lin}^{\rm II} & = \phantom{+}\sum_{n=0}^\infty \left\{ \nuii{n}
  \nLii{n} +   \nuiib{n} \nLiib{n} + \nuiic{n} \nLiic{n} \right \} \nonumber \\
&\quad + \sum_{n=1}^\infty \left\{ \nuv{n} \nLv{n} + \nuvb{n} \nLvb{n} +
  \nuvc{n}  \nLvc{n} + \nuvi{n} \nLvi{n} + \nuvib{n} \nLvib{n} \right\}\,.
\label{DeltalinII}
\end{align}
The commutation relations $[\w{}{\sigma}(z),\w{}{\sigma}(z')]=0$,
$[\wb{}{\sigma}(z'),\w{}{\sigma}(z)]=0$ yield the following system of equations.
\begin{subequations}
\label{consistencyII}
\begin{align}
-\nuii{n} - (n\!+\!1) \nuii{n+1} -\nuiib{n} -\tfr{n-1}{n} \nuvib{n-1} + \nuvib{n}
 &\,=0 & (n\ge 2) \\[0.1cm]
-\nuv{n} -(n\!+\!1) \nuv{n+1} -\nuvc{n} + (n\!+\!1) \nuvc{n+1} -\tfr{1}{n} \nuvib{n-1}
 &\,=0 
 & (n\ge 2) \\[0.1cm]
-\nuvb{n} -(n\!+\!1) \nuvb{n+1} + \tfr{1}{n} \nuvib{n-1} &\,=0  & (n\ge 2)
 \\[0.1cm] 
-\nuiic{n} -(n\!+\!1) \nuiic{n+1} -\nuvi{n} + (n\!+\!1) \nuiib{n+1} + \tfr{n-1}{n}
 \nuvib{n-1} &\,= 0 & (n\ge 1) \\[0.1cm]
 \nuvb{n} -(n\!+\!1) \nuvb{n+1} + \tfr{1}{n} \nuvi{n-1} &\, =0 & (n\ge 2) 
 \\[0.1cm]
-\nuii{1} -2\nuii{2} -\nuiib{1} + \nuvib{1} &\,=0\\[0.1cm]
-\nuv{1} -2\nuv{2} -\nuvc{1} +2\nuvc{2} &\,=0 \\[0.1cm]
-\nuvb{1} -2\nuvb{2} -\nuvi{1} + \nuvib{1}&\,=0 \\[0.1cm]
-\nuii{0} -\nuv{1} -\nuii{1} -\nuiib{0} + \nuvc{1} &\,=0 \\[0.1cm]
-\nuiic{0} -\nuvb{1} -\nuiic{1} + \nuiib{1} &\,=0 \\[0.1cm]
\nuiic{0} + \nuii{1} -\nuvb{1} &\,=0 \\[0.1cm]
2\nuii{2} -\nuvi{1} + \nuiic{1} &\,=0
\end{align}
\label{consII}
\end{subequations}
Again, we define some linear combinations of the Zimmermann coefficients,
\begin{subequations}
\label{vdefII}
\begin{align}
\qquad\qquad\qquad\qquad\qquad \vvi{n} &\,\equiv\, \half (\nuvi{n} - \nuvib{n} ) &(n\ge 1) \\[0.1cm]
\vii{n} &\,\equiv\, \nuii{n} -\tfr{n-1}{n} \vvi{n-1} + \vvi{n} &(n\ge 2) 
\\[0.1cm]
\vv{1} &\,\equiv -\half \nuii{0}-\half \nuiib{0}+\half\nuiic{0} &\\[0.1cm]
\vv{2} &\,\equiv\, \half \vv{1} -\half \nuv{1} &\\[0.1cm]
\vv{n+1} &\,\equiv\, \tfr{1}{n+1} \vv{n} -\tfr{1}{n+1} \nuv{n} -\tfr{1}{n(n+1)}
\vvi{n-1} & (n\ge 2) \\[0.1cm]
\vii{1} &\,\equiv\, \nuii{1} +\vvi{1} &\\[0.1cm]
\vii{0} &\,\equiv\, \nuii{0}+ \vv{1}\,. &
\end{align}
\end{subequations}
Equations (\ref{consII}) determine all Zimmermann coefficients in terms of
$\vii{n}$, $\vv{n}$ and $\vvi{n}$.

Calculation of the Weyl variation of $\Geff^{\rm lin \,\, II}$ shows that
the breaking can be completely absorbed by fixing the counterterm
coefficients 
\begin{equation}
\lambda_{\rm IIa}^{(n)}=\vii{n}\,, \qquad \lambda_{\rm
  IIb}^{(n)}=\vv{n}\,, \qquad \lambda_{\rm IIc}^{(n)}=\vvi{n}\,.
\label{linabsorbII}
\end{equation}
With the choice (\ref{linabsorbI}), (\ref{linabsorbII}), we have 
$S_{\rm lin}=0$
in (\ref{Sdecompose}).

\subsection*{The limit \boldmath $l=0$\unboldmath}
The Zimmermann coefficients of $0^{\rm th}$ order terms in $l$ are
related to the Zimmermann coefficients of the old minimal formalism
(cf. (\ref{Sexplicit}) as follows
\begin{gather}
\ui=\nui{0}\!,\qquad \uib=\nuib{0} \!, \nonumber \\
\uii=\nuii{0} + \nuiic{0} \!,\qquad \uiib=\nuiib{0}\!, \qquad
\ubox=-\nuiic{0}\,. \nonumber
\end{gather}
The counterterm coefficients to zeroth and first order in $l$ may be read
off from (\ref{vdefI}), (\ref{vdefII}) and the solution of (\ref{consistencyI}) and (\ref{consII}),
\begin{align}
\lambda_{\rm Ia}^{(0)} &\equiv \lambda_1 = \quar (\uib-\ui)&
\lambda_{\rm IIa}^{(0)} &\equiv \lambda_2 = \half (\uii-\uiib) \nonumber \\
\lambda_{\rm Ia}^{(1)} &= -\half (\uib+\ui)&
\lambda_{\rm IIa}^{(1)} &= \ubox \nonumber \\
&&
\lambda_{\rm IIb}^{(1)} &= -\half (\uii+\uiib)-\ubox\,. \nonumber
\end{align}
Again, we see that to 0$^{\rm th}$ order in $l$, the counterterm
coefficients (\ref{Rfix}) are reproduced,
while the remaining breaking terms of (\ref{Sexplicit}) are coupled to $l$.

This completes the proof of theorem \ref{thm61}.

\end{appendix}


\begin{thebibliography}{99}

\bibitem{PScurrent}T.E.~Clark, O.~Piguet, K.~Sibold, {\sl Supercurrents,
    Renormalization and Anomalies}, Nucl.~Phys. {\bf B 143} (1978) 445;\\
O.~Piguet, K.~Sibold, {\sl On the Renormalization of $N=1$ Rigid
  Supersymmetric Theories}, in: Supersymmetry and Supergtavity 1983, ed:
B.~Milewski, Proceedings of the XIXth Winter School and Workshop of
Theoretical Physics, Karpacz, Poland 1983.

\bibitem{PSBuch} O.~Piguet and K.~Sibold, {\sl Renormalized
    Supersymmetry. The Perturbation Theory of $N=1$ Supersymmetric Theories
    in Flat Space-Time}. 
Birkh\"auser, Boston 1986.

\bibitem{ERS1} J.~Erdmenger, C.~Rupp and K.~Sibold,  {\sl Conformal
    Transformation Properties of the Supercurrent in Four Dimensional
    Supersymmetric Theories}, Nucl.~Phys. {\bf B 530} (1998) 501,
    hep-th/9804053. 

\bibitem{ERS2} J.~Erdmenger, C.~Rupp, 
{\sl Geometrical superconformal anomalies},  hep-th/9809090.

\bibitem{ERS3} J.~Erdmenger, C.~Rupp, {\sl Superconformal Ward Identities for
  Green Functions with Multiple Supercurrent Insertions},
    Ann. of Phys. {\bf 276} (1999) 152-187,  hep-th/9811209. 

\bibitem{ERS4}  J.~Erdmenger, C.~Rupp and K.~Sibold, {\sl Superconformal
    Transformation Properties of the Supercurrent II: Abelian Gauge
    Theories},  hep-th/9907169, to be published in Nucl.~Phys.~{\bf B}.

\bibitem{newmin} S.J.~Gates, M.~Ro$\breve{\rm c}$ek, W.~Siegel, {\sl
    Solution to Constraints for $n=0$ Supergravity}, Nucl.~Phys. {\bf B
    198} (1982) 113. \\
  G.~Girardi, R.~Grimm, {\sl Superspace Geometry and the Minimal, Non
    Minimal, and New Minimal Supergravity Multiplets}, Z.~Phys. {\bf C 26}
  (1984) 123.

\bibitem{Grisaru} S.J. Gates, M.T. Grisaru, M. Ro$\breve{\rm c}$ek and
  W. Siegel, 
{\sl Superspace}. The Benjamin/Cummings Publishing Company, Inc. 1983.

\bibitem{Buchbinder}I.L.~Buchbinder and S.M.~Kuzenko,  
{\sl Ideas and Methods of Supersymmetry and Supergravity}. 
Institute of Physics Publishing, Bristol and Philadelphia 1995.



\bibitem{PW}O.~Piguet, S.~Wolf, {\sl The Supercurrent Trace Indentities of
    the $N=1$, $D=4$ Super Yang-Mills Theory in the Wess-Zumino Gauge},
  JHEP 9804 (1998) 1.

\bibitem{bphz} J.H.~Lowenstein, {\sl BPHZ renormalization}, in: Renormalization
  Theory, eds. G.~Velo and A.S.~Wightman, Erice lectures 1976; \newline
  O.~Piguet and S.~Sorella, {\sl Algebraic Renormalization}. Springer Verlag,
  Berlin 1995; \newline
  K.~Sibold, {\sl St\"orungstheoretische Renormierung. Quantisierung von
  Eichtheorien}. Vorlesungsskript TU M\"unchen 1993.

\bibitem{composite}
  W.~Zimmermann, {\sl Composite Operators in the Perturbation Theory of
  Renormalizable Interactions}, Ann. of Phys. {\bf 77} (1973) 536;\newline
  W.~Zimmermann, {\sl Normal Products and the Short Distance Expansion in
  the Perturbation Theory of Renormalizable Interactions}, Ann. of
  Phys. {\bf 77} (1973) 570.

\bibitem{KS1} E.~Kraus and K.~Sibold, {\sl Conformal Transformation
    Properties of the Energy Momentum Tensor in
Four Dimensions}, Nucl.~Phys. {\bf B 372} (1992) 113.


\bibitem{Osborn} H.~Osborn, {\sl $N=1$ Superconformal Symmetry in 
  Four Dimensional Quantum Field Theory}, hep-th/9808041.

\end{thebibliography}
\end{document}